\documentclass[%
reprint,
superscriptaddress,
bibnotes,
amsmath,amssymb,
aps,
prl,
longbibliography,
]{revtex4-2}

\usepackage{graphicx}% Include figure files
\usepackage{dcolumn}% Align table columns on decimal point
\usepackage{bm}% bold math
\usepackage{chemformula}
\usepackage{hyperref}% add hypertext capabilities
\begin{document}

%\preprint{APS/123-QED}

\title{Persistent spin dynamics in magnetically ordered honeycomb cobalt oxides}

\author{Ping~Miao}
\email{miaoping@ihep.ac.cn}
\affiliation{Institute of High-Energy Physics, Chinese Academy of Sciences, Beijing 100049, China}
\affiliation{Spallation Neutron Source Science Center, Dongguan 523803, China}
\author{Xianghong~Jin}
\affiliation{International Center for Quantum Materials, School of Physics, Peking University, Beijing 100871, China}
\author{Weiliang~Yao}
\affiliation{International Center for Quantum Materials, School of Physics, Peking University, Beijing 100871, China}
\author{Yue~Chen}
\affiliation{International Center for Quantum Materials, School of Physics, Peking University, Beijing 100871, China}
\author{Akihiro~Koda}
\affiliation{Institute of Materials Structure Science, High Energy Accelerator Research Organization (KEK), Tokai, Ibaraki 319-1106, Japan}
\affiliation{Department of Materials Structure Science, Sokendai (The Graduate University for Advanced Studies), Tokai, Ibaraki 319-1106, Japan}
\author{Zhenhong~Tan}
\affiliation{Institute of High-Energy Physics, Chinese Academy of Sciences, Beijing 100049, China}
\affiliation{Spallation Neutron Source Science Center, Dongguan 523803, China}
\author{Wu~Xie}
\affiliation{Institute of High-Energy Physics, Chinese Academy of Sciences, Beijing 100049, China}
\affiliation{Spallation Neutron Source Science Center, Dongguan 523803, China}
\author{Wenhai~Ji}
\affiliation{Institute of High-Energy Physics, Chinese Academy of Sciences, Beijing 100049, China}
\affiliation{Spallation Neutron Source Science Center, Dongguan 523803, China}
\author{Takashi~Kamiyama}
\affiliation{Institute of High-Energy Physics, Chinese Academy of Sciences, Beijing 100049, China}
\affiliation{Spallation Neutron Source Science Center, Dongguan 523803, China}
\author{Yuan~Li}
\email{yuan.li@pku.edu.cn}
\affiliation{International Center for Quantum Materials, School of Physics, Peking University, Beijing 100871, China}
\date{\today}% It is always \today, today,

\begin{abstract}
In the quest to find quantum spin liquids, layered cobalt oxides \ch{Na_2Co_2TeO_6} and \ch{Na_3Co_2SbO_6} have been proposed as promising candidates for approximating the Kitaev honeycomb model. Yet, their suitability has been thrown into question due to observed long-range magnetic order at low temperatures and indications of easy-plane, rather than Kitaev-type, spin anisotropy. Here we use muon spin relaxation to reveal an unexpected picture: contrary to the anticipated static nature of the long-range order, the systems show prevalent spin dynamics with spatially uneven distribution and varied correlation times. This underlines that the magnetic ground states cannot be solely described by the long-range order, suggesting a significant role of quantum fluctuations. Our findings not only shed new light on the complex physics of these systems but also underscore the need for a refined approach in the search for realizable quantum spin liquids.

\end{abstract}

\maketitle

Materials that might realize the Kitaev model for quantum spin liquids \cite{KitaevAP2006} have attracted considerable interest \cite{TakagiNRP2019,TrebstPRRSP2022}. The model represents a distinct route to magnetic frustration on the honeycomb lattice, where each spin cannot simultaneously satisfy the bond-dependent and mutually orthogonal Ising interactions with its three neighbors. Realizing these interactions in crystalline materials involves a delicate balance \cite{JackeliPRL2009,ChaloupkaPRL2010,PlumbPRB2014,LiuPRB2018,SanoPRB2018,MotomeJPCM2020} among electron interaction, crystal-field splitting, and spin-orbit coupling. In particular, the desired Kitaev-type three-dimensional (3D) interaction anisotropy is at variance with the two-dimensional (2D) geometry of the honeycomb lattice. To best approximate the Kitaev model, the crystal-field and hopping parameters need to be locally restricted to the magnetic ions and their ligands, without significantly involving more distant atoms that reflect the lattice geometry.

Recently, cobalt oxides with a layered honeycomb structure, including \ch{Na_2Co_2TeO_6} (NCTO), \ch{Na_3Co_2SbO_6} (NCSO), and \ch{BaCo_2(AsO_4)_2} (BCAO), have been intensely studied for their potential to realize the Kitaev model \cite{LiuPRB2018,SanoPRB2018,LiuPRL2020,YanPRM2019,YaoPRB2020,ZhongSA2020,SongvilayPRB2020,LinNC2021,KimJPCM2022b,KimJPCM2022}. Compared to the $4d$ and $5d$ counterparts such as in $\alpha$-\ch{RuCl_3} \cite{PlumbPRB2014} and \ch{Na_2IrO_3} \cite{ChaloupkaPRL2010}, respectively, the Co$^{2+}$ $3d$ orbitals in these compounds are expected to be more localized. However, whether this localization is sufficient to suppress direct (i.e., not via ligands) hopping between nearest neighbors and farther-neighbor hopping remains a point of contention \cite{DasPRB2021,KimJPCM2022,MaksimovPRB2022,WinterJOP2022,PandeyPRB2022,LiuPRB2023,HalloranPNAS2023}. While farther-neighbor hopping might be treated as a perturbation to the nearest-neighbor (Kitaev) model, which can already influence important characteristics of the magnetic order and excitations \cite{YaoPRL2022,KrugerArxiv2022,PandeyPRB2022,WangArxiv2023}, direct hopping is believed to lead to a strong departure from the Kitaev model concerning the interaction anisotropy \cite{DasPRB2021,KimJPCM2022,MaksimovPRB2022,WinterJOP2022,LiuPRB2023,HalloranPNAS2023}. As most of the cobalt oxides do develop long-range magnetic order at low temperatures, such departures from the ideal Kitaev model cannot be neglected.

A crucial aspect related to the direct nearest-neighbor hopping, when combined with crystal-field and spin-orbit effects, is whether the cobalt oxides align more with an easy-plane (XXZ) model than the Kitaev model. Ab initio calculations suggest that this may indeed be the case for NCTO and BCAO  \cite{DasPRB2021,MaksimovPRB2022,WinterJOP2022,LiuPRB2023}. Yet, even for NCSO, which is considered more compatible with a Kitaev-like theoretical starting point \cite{LiuPRB2023,VavilovaPRB2023,KangPRB2023}, recent studies present evidence supporting XXZ-like anisotropy \cite{GuArxiv2023}. Notably, XXZ- and Kitaev-like models are very different in their magnetic frustration properties: A bond-independent XXZ model is bipartite on the honeycomb lattice, such that it needs to be supplemented by farther-neighbor interactions in order to be frustrated \cite{DasPRB2021,MaksimovPRB2022,HalloranPNAS2023,WinterJOP2022,NairPRB2018}. Even in the fully bond-dependent Ising limit, the frustration is still arguably weaker than in the Kitaev model, because the model has no longer three mutually orthogonal spin anisotropy axes (instead, they are 120$^\circ$ apart). It is thus important to search for signs of magnetic quantum fluctuations in the cobalt oxides to evaluate their potential for manifesting novel quantum states of matter.

\begin{figure}[!ht]
\includegraphics[width=3.37in]{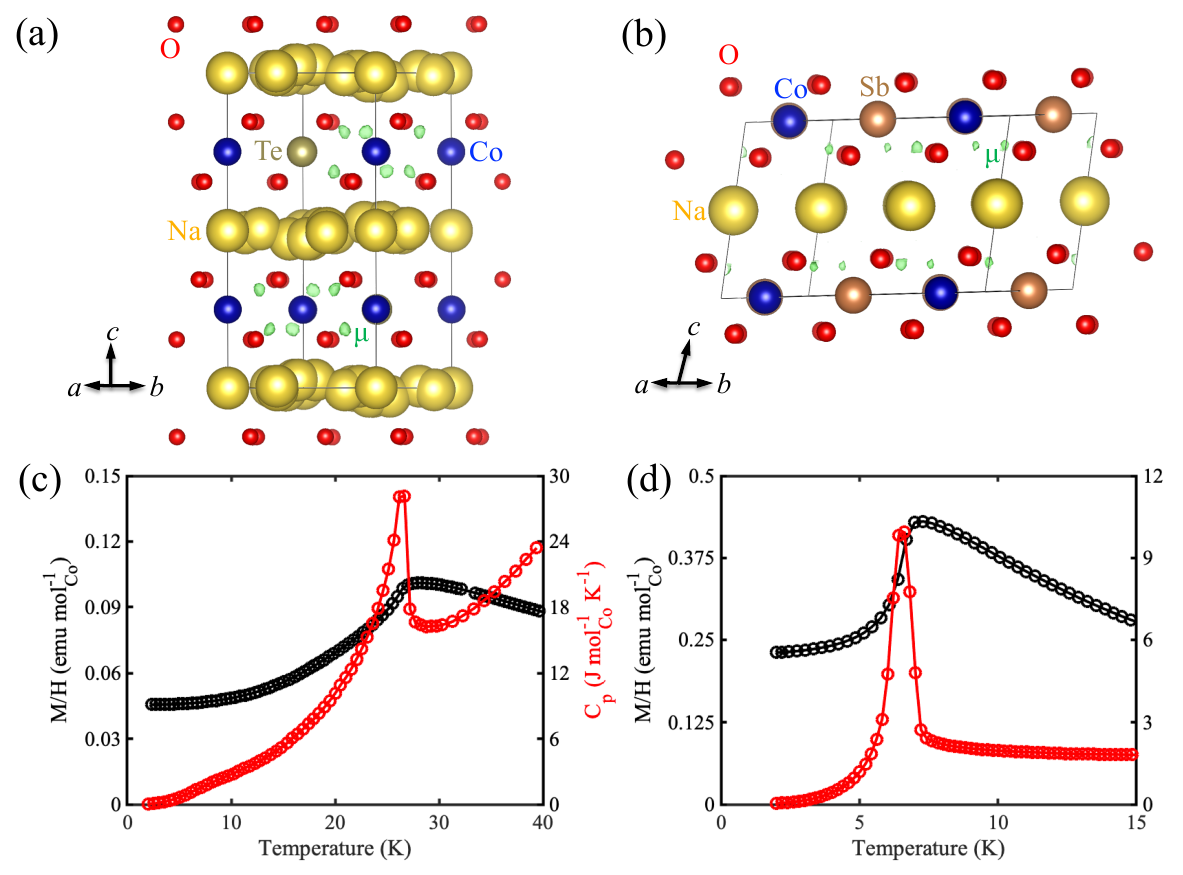}
\caption{(a-b) Crystal structures of \ch{Na_2Co_2TeO_6} and \ch{Na_3Co_2SbO_6}, respectively, along with calculated muon stoppage sites indicated by small spheres near the cobalt atoms. (c-d) Magnetic susceptibility of \ch{Na_2Co_2TeO_6} and \ch{Na_3Co_2SbO_6}, measured on single crystals with in-plane magnetic fields of 5000 Oe and 1000 Oe, respectively. The data are plotted together with specific heat data measured in zero field to demonstrate $T_\mathrm{N}$ and the homogeneity of the samples.}
\label{fig1}
\end{figure}

In this paper, we present a comprehensive muon ($\mu^+$) spin relaxation ($\mu$SR) study of NCTO and NCSO, conducted from the paramagnetic states to deep inside their magnetically ordered states. By varying temperatures and applying longitudinal magnetic fields, we were able to distinguish between muon-spin relaxation caused by static and dynamic magnetic moments. In both materials, we discovered that the dynamic relaxation rate peaks at a temperature substantially below the ordering temperature ($T_\mathrm{N}$), displaying a stretched-exponential behavior indicative of a glass-like distribution of relaxation times. Crucially, we found that a considerable fraction of the spins maintain their dynamic nature even at temperatures that are an order of magnitude below $T_\mathrm{N}$. These observations provide compelling evidence for significant quantum fluctuations in both systems, underscoring their highly frustrated nature. The persistent spin dynamics could also impose constraints on the nature of the magnetic ground states.

Figure \ref{fig1}(a-b) compares the crystal structures of NCTO and NCSO. The honeycomb cobalt layers are formed by edge-sharing CoO$_6$ octahedra. They are separated by sodium atoms and have weak inter-layer magnetic coupling \cite{ChenPRB2021,YaoPRL2022}. While this renders the difference in the inter-layer stacking of the two systems (hexagonal in NCTO and monoclinic in NCSO) seemingly unimportant, the monoclinic distortion of NCSO affects the Co-O bonding geometry and produces dramatic in-plane magnetic anisotropy \cite{LiPRX2022}. Both systems show large in-plane versus out-of-plane magnetization anisotropy \cite{YaoPRB2020,YanPRM2019,LiPRX2022}, which can be attributed to either global XXZ-like easy-plane anisotropy, or bond-dependent Kitaev-like anisotropy supplemented by off-diagonal coupling \cite{JanssenPRB2017}. Their transition temperatures $T_\mathrm{N}$ differ by about a factor of four [Fig.~\ref{fig1}(c-d)], and it is believed that NCSO is closer to a ferromagnetic instability \cite{LiuPRL2020,LiPRX2022}.

\begin{figure}[!ht]
\includegraphics[width=3.37in]{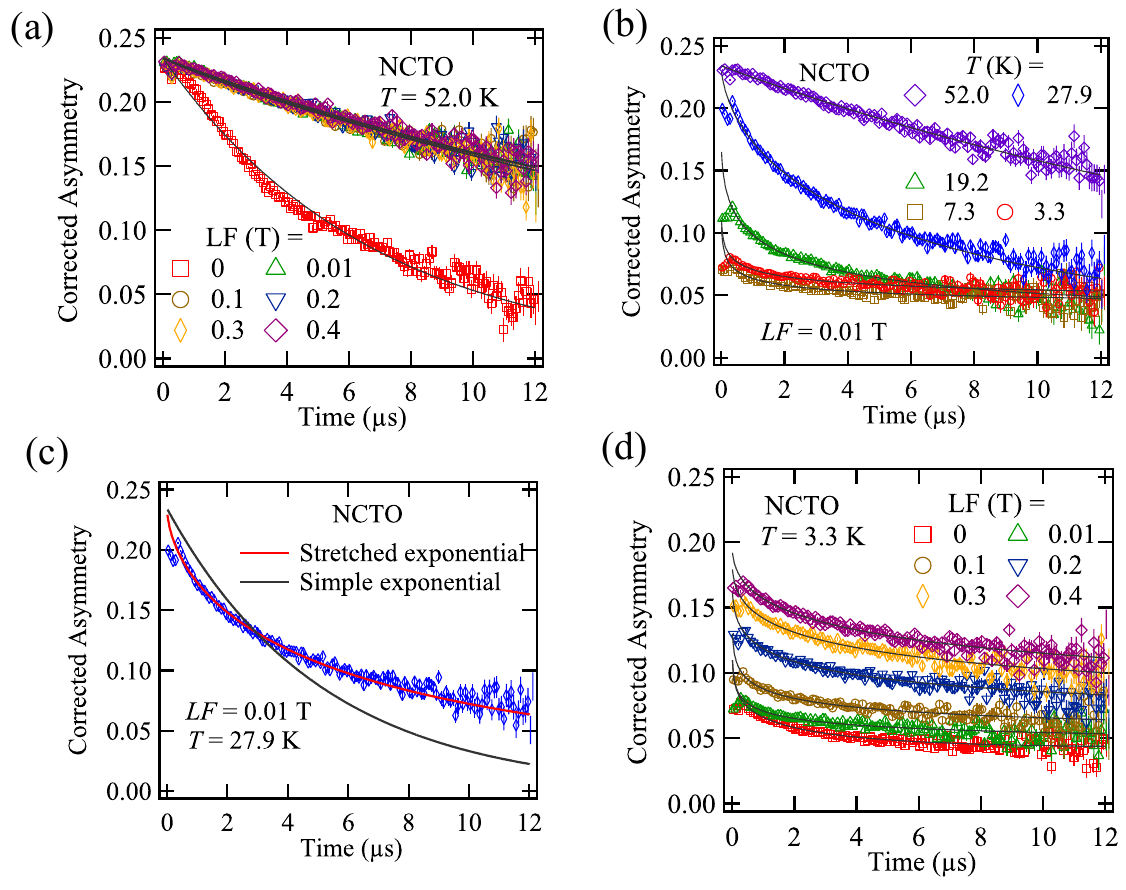}
\caption{(a) $\mu$SR spectra of NCTO measured far above $T_\mathrm{N}$. A longitudinal field (LF) of 0.01 T decouples the nuclear dipolar fields. (b) Near-zero-field $\mu$SR spectra measured at selected temperatures. (c) Demonstration of stretched exponential fitting of the data obtained in near-zero field and just above $T_\mathrm{N}$. (d) Lowest-temperature spectra under various LFs, revealing a dual static and dynamic nature of the relaxation.}
\label{fig2}
\end{figure}

$\mu$SR is a powerful probe of both static and dynamic magnetization in solids. In both NCTO and NCSO, implanted muons are expected to stop not far from the cobalt layers [Fig.~\ref{fig1}(a-b)], which makes them very sensitive to internal magnetic fields generated by the cobalt atoms. Our $\mu$SR experiments were performed on the ARTEMIS beam line at the Japan Proton Accelerator Research Complex (J-PARC), which utilizes a double-pulse high-count-rate method. Polycrystalline NCTO and NCSO were pressed into pellets of about 5 mm in thickness and 25 mm in diameter. NCTO was mounted in a helium-4 cryostat while NCSO was measured in both helium-4 and helium-3 cryostats. All the $\mu$SR data of NCSO discussed in the paper were from the helium-3 cryostat except for that in Fig.~\ref{figS3}(a) \cite{SM}. The base temperatures of helium-4 and helium-3 cryostats are about 3.3 K and 0.3 K, which are about one order of magnitude below the respective $T_\mathrm{N}$ of the systems. We performed our measurements under a variety of longitudinal magnetic fields (LFs, up to $H=0.4$~Tesla) parallel to the initial $\mu^+$ spin direction, and analyzed the data using the software suite \textit{musrfit} \cite{SuterPP2012}. $\mu$SR spectra are represented by the corrected asymmetry of the muon-decay counts as a function of delay time after implantation.

Figure \ref{fig2}(a) displays the $\mu$SR spectra of NCTO obtained at 52 K. As the temperature is far above $T_\mathrm{N}=$ 26.7 K, the observed relaxation is not caused by static internal fields related to the magnetic order. The significant difference between the zero-field (ZF) and the LF = 0.01 T spectra indicates that the former is affected by cobalt nuclear magnetic moments, which the weak LF is sufficient to decouple. As such a weak LF is not expected to be comparable to internal magnetic fields generated by the Co electronic magnetic moments (over 2 $\mu_\mathrm{B}$ in NCTO \cite{YaoPRB2020}), we continue to display in Fig.~\ref{fig2}(b) the spectra obtained with LF = 0.01 T at lower temperatures.

For a quantitative analysis, we fit the spectra to a sum of dynamic and static components:
\begin{equation}
\begin{split}
\label{eq1}
A_0 P_\mathrm{z}(H, t) =  &A_{bg} + A_1 \exp[-(\Lambda t)^\beta] \\
&+ A_2 [\alpha(H) + (1-\alpha(H)) \cos(\gamma_\mu B t)],
\end{split}
\end{equation}
where $A_0$ is the initial asymmetry and $P_\mathrm{z}(H, t)$ the normalized polarization function. $A_1$ and $A_2$ are magnitudes of contributions from the dynamic and static phases, respectively, and $A_{bg}$ denotes the background asymmetry. $A_{bg}$ is negligible for NCTO and determined to be 0.025 for NCSO, as shown by the ZF spectra at 26.6 K and 13.3 K in Figs.~\ref{figS1} and \ref{figS3} in \cite{SM}, respectively. The dynamic contribution is best described by a stretched-exponential relaxation function, where $\Lambda$ is the dynamic spin-lattice-relaxation rate and $\beta$ the exponent, the physical meaning of which will be discussed later. $A_2$ is zero above $T_\mathrm{N}$ and increases (at the cost of decreasing $A_1$) below $T_\mathrm{N}$. $\alpha$ accounts for the effect of LFs on the relaxation caused by the magnetic order. Because of powder average, $\alpha$ equals $1/3$ at LF = 0.01 T (or ZF) and gradually increases with increasing LFs. This is manifested by the fact that the data in Fig.~\ref{fig2}(d) for LF above 0.01 T are approximately offset from one another along the vertical direction. The oscillation term $\cos(\gamma_\mu B t)$ is assumed to average to zero in our measurements due to the limited time resolution. Additional temperature- and LF-dependent data of NCTO are displayed in Fig.~\ref{figS1} \cite{SM}, and similarly obtained data as in Fig.~\ref{fig2} but for NCSO are displayed in Fig.~\ref{figS2} \cite{SM}.

\begin{figure}[!ht]
\includegraphics[width=2.4in]{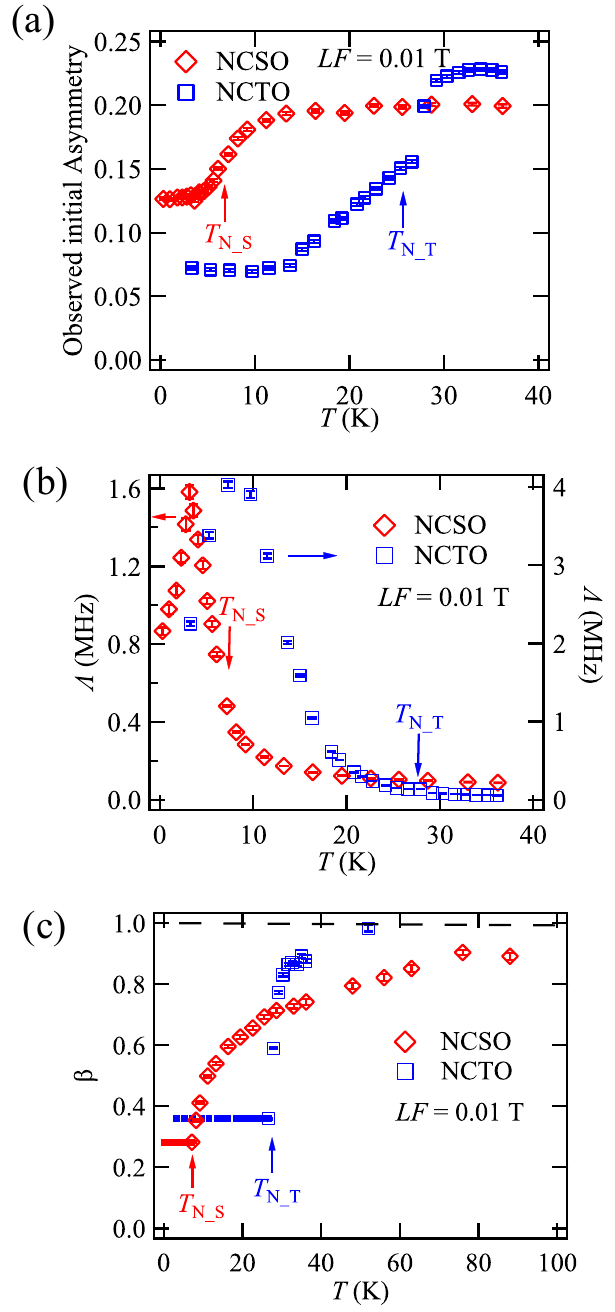}
\caption{(a) Muon-spin asymmetry in the nominal $t=0$ limit measured by our relatively poor time resolution. (b) Muon relaxation rate as a function of temperature. The relaxation is primarily caused by dynamic internal fields because longitudinal fields up to 0.4 T have little effect on it [Fig.~\ref{fig2}(d). (c) Stretched exponent used in our fits to the time spectra (see text). Below $T_\mathrm{N}$, the data are consistent with a stable exponent and are therefore fit to a common value. Arrows indicate $T_\mathrm{N}$ of the two systems.}
\label{fig3}
\end{figure}

Figure \ref{fig3} summarizes our findings from the fits for both NCTO and NCSO. At LF = 0.01 T, we observe a clear decrease in the initial asymmetry (corrected asymmetry at $t=0$) upon cooling below $T_\mathrm{N}$ as the long-range magnetic order forms [Fig.~\ref{fig3}(a)]. This can be directly seen from Fig.~\ref{fig2}(b), where each of the five spectra starts from a different initial value. We attribute this behavior to the fact that the internal magnetic fields associated with the static phase ($A_2$ term in Eq.~\ref{eq1}) are strong, making the asymmetry oscillate too rapidly to be resolved by our time resolution. Our understanding about this static-phase contribution is corroborated by the spectra under LFs at 3.3 K [Fig.~\ref{fig2}(d)], where the initial asymmetry increases with increasing LFs. Especially for NCTO, the observed initial asymmetry does not behave like a typical order parameter below $T_\mathrm{N}$. Instead, it undergoes a gradual, nearly linear decrease with cooling at LF = 0.01 T, and levels off below about 16 K [Fig.~\ref{fig3}(a)]. This indicates that the system somehow resists to develop conventional long-range order, possibly due to magnetic frustration and the existence of closely competing ground states.

After the initial decrease in asymmetry, the spectra in  [Fig.~\ref{fig2}(b)] show gradual damping over time, which is observed even under LF = 0.4 T at 3.3 K [Fig.~\ref{fig2}(d)], indicating that a dynamic phase coexists with the static phase. This dynamic contribution is characterized by the relaxation rate $\Lambda$ and the stretched exponent $\beta$ ($A_1$ term in Eq.~\ref{eq1}). It is clear from Fig.~\ref{fig2}(c) that a simple exponential function, i.e., with $\beta=1$, would fail to describe the data obtained near $T_\mathrm{N}$. Instead, we find that $\beta$ decreases from 1 steeply to about 1/3 as the sample is cooled towards $T_\mathrm{N}$, and stays that way below $T_\mathrm{N}$ [Fig.~\ref{fig3}(c)]. Significant deviations from a simple exponential relaxation function usually signify a broad distribution of spatially inhomogeneous correlation times, which is commonly observed in glassy magnets \cite{UemuraPRB1985,HeffnerPRL1996,MiaoAPL2019}. Indeed, $\beta= 1/3$ has been reported in the metallic spin glass AgMn \cite{CampbellPRL1994,KerenPRL1996}. Since NCTO and NCSO are chemically ordered systems, we attribute our observation of stretched exponents to spin-glass-like behaviors which have previously been observed in geometrically frustrated antiferromagnets \cite{DunsigerJAP1996,GardnerPRL1999,DeyPRB2023}. This understanding is further supported by the unusual temperature evolution of the relaxation rate $\Lambda$, which serves as a one-parameter abstraction of the relaxation process. Namely, we find that $\Lambda$ reaches its maximum only at a temperature significantly lower than $T_\mathrm{N}$, which is inconsistent with a conventional second-order phase transition at $T_\mathrm{N}$  \cite{HeffnerPRL1996, Gubbens1994}. The data suggest that substantial spin fluctuations continue to exist below $T_\mathrm{N}$ until a much lower temperature is reached. This is consistent with previous inelastic neutron scattering results on NCTO \cite{ChenPRB2021}, where spectrally well-defined spin waves were only observed far below $T_\mathrm{N}$.

\begin{figure}[!t]
\includegraphics[width=3.37in]{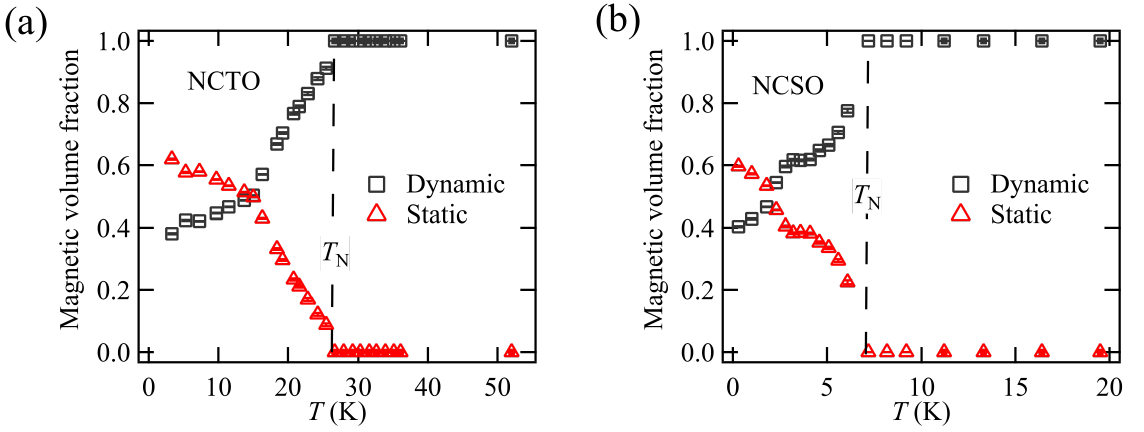}
\caption{Volume fraction of dynamic ($A_1 / (A_1+A_2)$) and static phases ($A_2 / (A_1+A_2)$) as determined from the fit values of $A_1$ and $A_2$ according to Eq.~\ref{eq1}. $A_2=0$ has been assumed in the fits above $T_\mathrm{N}$.}
\label{fig4}
\end{figure}

Taken together, our $\mu$SR data indicate that the magnetic orders in NCTO and NCSO are distinct from conventional long-range magnetic order \cite{MiaoPRB2021}, as they (1) are not accompanied by an anomaly of $\Lambda$  at $T_\mathrm{N}$ [Fig.~\ref{fig3}(b)], (2) exhibit a ``glassy'' growth of the internal fields below $T_\mathrm{N}$ [Fig.~\ref{fig3}(a)], and (3) have wide distribution of correlation times even in the well-ordered state [Fig.~\ref{fig3}(c)]. To go one step further, we can use $A_1$ and $A_2$ in Eq.~\ref{eq1} to estimate the volume fractions that are dynamic and static, respectively, at a given temperature [Fig.~\ref{fig4}]. In line with the unconventional nature of the orders, the result reveals that a significant fraction of the samples remains dynamic down to the lowest temperature. In fact, despite the different $T_\mathrm{N}$, the two systems look remarkably similar. In the case of NCTO, it is known that the spin excitation spectrum exhibits a gap of over 1 meV at 3.3 K \cite{ChenPRB2021,LinNC2021,YaoPRL2022}, which makes thermal fluctuations unable to cause significant spin dynamics in the system. Because NCSO exhibits even stronger spin anisotropy than NCTO \cite{LiPRX2022}, it is likely that in the well-ordered state the system has an even larger excitation gap than NCTO. However, our data show that the dynamic volume fraction in NCSO is as large as 40\% even at the very low temperature of 0.3 K. We therefore believe that the observed spin dynamics are caused by quantum rather than thermal fluctuations.

The presence of significant spin dynamics in both NCTO and NCSO deep in the ordered state suggests that the systems may be driven into a quantum paramagnetic state if the semi-classical order can be suppressed. We further speculate that the dynamic phase volumes, up to about 40\% in both systems at the lowest temperature, may corroborate recent proposals of multi-$\mathbf{q}$ magnetic orders in the two systems \cite{ChenPRB2021,YaoPRR2023,GuArxiv2023}. Specifically, if the spin anisotropy of these cobalt oxides is indeed XXZ- rather than Kitaev-like \cite{DasPRB2021,KimJPCM2022,MaksimovPRB2022,WinterJOP2022,LiuPRB2023,HalloranPNAS2023}, meaning that the ordered moments all lie parallel to the $ab$ plane \cite{GuArxiv2023}, the proposed multi-$\mathbf{q}$ orders would feature 1/4 of the cobalt sites being (classically)  ``spinless'' (in spite of large ordered moment size measured in diffraction \cite{LefrancoisPRB2016,BeraPRB2017}) in NCTO \cite{ChenPRB2021} and half of the sites having smaller classical moments than the other half \cite{GuArxiv2023}. Such scenarios are in good qualitative agreement with the dynamic volume fractions we observe.

To conclude, we have discovered persistent and spatially inhomogeneous spin dynamics deep inside the magnetically ordered states of two layered honeycomb cobalt oxides. Our findings indicate that the systems are highly frustrated despite their formation of long-range order, and imply that they continue to hold promise for realizing novel quantum states of matter if the semi-classical aspect of the ground states can be further suppressed.

\begin{acknowledgments}
We are grateful for technical support for calculation of muon stoppage sites by Dr. Hua Li. The work was supported by the National Basic Research Program of China (Grant No. 2021YFA1401900), the NSF of China (Grant Nos. 12061131004, 11888101, 12005243 and 22205239), and the Guangdong Basic and Applied Basic Research Foundation (Grant No. 2022B1515120014).  The $\mu$SR experiments were performed at the MLF, J-PARC, Japan, under a user program (No. 2019B0201 and 2022A0141).

\end{acknowledgments}

\bibliography{NCTO_NCSO_muSR_ref}

\begin{thebibliography}{48}%
\makeatletter
\providecommand \@ifxundefined [1]{%
 \@ifx{#1\undefined}
}%
\providecommand \@ifnum [1]{%
 \ifnum #1\expandafter \@firstoftwo
 \else \expandafter \@secondoftwo
 \fi
}%
\providecommand \@ifx [1]{%
 \ifx #1\expandafter \@firstoftwo
 \else \expandafter \@secondoftwo
 \fi
}%
\providecommand \natexlab [1]{#1}%
\providecommand \enquote  [1]{``#1''}%
\providecommand \bibnamefont  [1]{#1}%
\providecommand \bibfnamefont [1]{#1}%
\providecommand \citenamefont [1]{#1}%
\providecommand \href@noop [0]{\@secondoftwo}%
\providecommand \href [0]{\begingroup \@sanitize@url \@href}%
\providecommand \@href[1]{\@@startlink{#1}\@@href}%
\providecommand \@@href[1]{\endgroup#1\@@endlink}%
\providecommand \@sanitize@url [0]{\catcode `\\12\catcode `\$12\catcode
  `\&12\catcode `\#12\catcode `\^12\catcode `\_12\catcode `\%12\relax}%
\providecommand \@@startlink[1]{}%
\providecommand \@@endlink[0]{}%
\providecommand \url  [0]{\begingroup\@sanitize@url \@url }%
\providecommand \@url [1]{\endgroup\@href {#1}{\urlprefix }}%
\providecommand \urlprefix  [0]{URL }%
\providecommand \Eprint [0]{\href }%
\providecommand \doibase [0]{https://doi.org/}%
\providecommand \selectlanguage [0]{\@gobble}%
\providecommand \bibinfo  [0]{\@secondoftwo}%
\providecommand \bibfield  [0]{\@secondoftwo}%
\providecommand \translation [1]{[#1]}%
\providecommand \BibitemOpen [0]{}%
\providecommand \bibitemStop [0]{}%
\providecommand \bibitemNoStop [0]{.\EOS\space}%
\providecommand \EOS [0]{\spacefactor3000\relax}%
\providecommand \BibitemShut  [1]{\csname bibitem#1\endcsname}%
\let\auto@bib@innerbib\@empty
%</preamble>
\bibitem [{\citenamefont {Kitaev}(2006)}]{KitaevAP2006}%
  \BibitemOpen
  \bibfield  {author} {\bibinfo {author} {\bibfnamefont {A.}~\bibnamefont
  {Kitaev}},\ }\bibfield  {title} {\bibinfo {title} {Anyons in an exactly
  solved model and beyond},\ }\href
  {https://doi.org/https://doi.org/10.1016/j.aop.2005.10.005} {\bibfield
  {journal} {\bibinfo  {journal} {Ann. Phys.}\ }\textbf {\bibinfo {volume}
  {321}},\ \bibinfo {pages} {2 } (\bibinfo {year} {2006})}\BibitemShut
  {NoStop}%
\bibitem [{\citenamefont {Takagi}\ \emph {et~al.}(2019)\citenamefont {Takagi},
  \citenamefont {Takayama}, \citenamefont {Jackeli}, \citenamefont
  {Khaliullin},\ and\ \citenamefont {Nagler}}]{TakagiNRP2019}%
  \BibitemOpen
  \bibfield  {author} {\bibinfo {author} {\bibfnamefont {H.}~\bibnamefont
  {Takagi}}, \bibinfo {author} {\bibfnamefont {T.}~\bibnamefont {Takayama}},
  \bibinfo {author} {\bibfnamefont {G.}~\bibnamefont {Jackeli}}, \bibinfo
  {author} {\bibfnamefont {G.}~\bibnamefont {Khaliullin}},\ and\ \bibinfo
  {author} {\bibfnamefont {S.~E.}\ \bibnamefont {Nagler}},\ }\bibfield  {title}
  {\bibinfo {title} {{Concept and realization of Kitaev quantum spin
  liquids}},\ }\href {https://doi.org/10.1038/s42254-019-0038-2} {\bibfield
  {journal} {\bibinfo  {journal} {Nat. Rev. Phys.}\ }\textbf {\bibinfo {volume}
  {1}},\ \bibinfo {pages} {264} (\bibinfo {year} {2019})}\BibitemShut {NoStop}%
\bibitem [{\citenamefont {Trebst}\ and\ \citenamefont
  {Hickey}(2022)}]{TrebstPRRSP2022}%
  \BibitemOpen
  \bibfield  {author} {\bibinfo {author} {\bibfnamefont {S.}~\bibnamefont
  {Trebst}}\ and\ \bibinfo {author} {\bibfnamefont {C.}~\bibnamefont
  {Hickey}},\ }\bibfield  {title} {\bibinfo {title} {Kitaev materials},\ }\href
  {https://doi.org/https://doi.org/10.1016/j.physrep.2021.11.003} {\bibfield
  {journal} {\bibinfo  {journal} {Physics Reports}\ }\textbf {\bibinfo {volume}
  {950}},\ \bibinfo {pages} {1} (\bibinfo {year} {2022})}\BibitemShut {NoStop}%
\bibitem [{\citenamefont {Jackeli}\ and\ \citenamefont
  {Khaliullin}(2009)}]{JackeliPRL2009}%
  \BibitemOpen
  \bibfield  {author} {\bibinfo {author} {\bibfnamefont {G.}~\bibnamefont
  {Jackeli}}\ and\ \bibinfo {author} {\bibfnamefont {G.}~\bibnamefont
  {Khaliullin}},\ }\bibfield  {title} {\bibinfo {title} {{Mott Insulators in
  the Strong Spin-Orbit Coupling Limit: From Heisenberg to a Quantum Compass
  and Kitaev Models}},\ }\href {https://doi.org/10.1103/PhysRevLett.102.017205}
  {\bibfield  {journal} {\bibinfo  {journal} {Phys. Rev. Lett.}\ }\textbf
  {\bibinfo {volume} {102}},\ \bibinfo {pages} {017205} (\bibinfo {year}
  {2009})}\BibitemShut {NoStop}%
\bibitem [{\citenamefont {Chaloupka}\ \emph {et~al.}(2010)\citenamefont
  {Chaloupka}, \citenamefont {Jackeli},\ and\ \citenamefont
  {Khaliullin}}]{ChaloupkaPRL2010}%
  \BibitemOpen
  \bibfield  {author} {\bibinfo {author} {\bibfnamefont {J.}~\bibnamefont
  {Chaloupka}}, \bibinfo {author} {\bibfnamefont {G.}~\bibnamefont {Jackeli}},\
  and\ \bibinfo {author} {\bibfnamefont {G.}~\bibnamefont {Khaliullin}},\
  }\bibfield  {title} {\bibinfo {title} {{Kitaev-Heisenberg Model on a
  Honeycomb Lattice: Possible Exotic Phases in Iridium Oxides
  ${A}_{2}{\mathrm{IrO}}_{3}$}},\ }\href
  {https://doi.org/10.1103/PhysRevLett.105.027204} {\bibfield  {journal}
  {\bibinfo  {journal} {Phys. Rev. Lett.}\ }\textbf {\bibinfo {volume} {105}},\
  \bibinfo {pages} {027204} (\bibinfo {year} {2010})}\BibitemShut {NoStop}%
\bibitem [{\citenamefont {Plumb}\ \emph {et~al.}(2014)\citenamefont {Plumb},
  \citenamefont {Clancy}, \citenamefont {Sandilands}, \citenamefont {Shankar},
  \citenamefont {Hu}, \citenamefont {Burch}, \citenamefont {Kee},\ and\
  \citenamefont {Kim}}]{PlumbPRB2014}%
  \BibitemOpen
  \bibfield  {author} {\bibinfo {author} {\bibfnamefont {K.~W.}\ \bibnamefont
  {Plumb}}, \bibinfo {author} {\bibfnamefont {J.~P.}\ \bibnamefont {Clancy}},
  \bibinfo {author} {\bibfnamefont {L.~J.}\ \bibnamefont {Sandilands}},
  \bibinfo {author} {\bibfnamefont {V.~V.}\ \bibnamefont {Shankar}}, \bibinfo
  {author} {\bibfnamefont {Y.~F.}\ \bibnamefont {Hu}}, \bibinfo {author}
  {\bibfnamefont {K.~S.}\ \bibnamefont {Burch}}, \bibinfo {author}
  {\bibfnamefont {H.-Y.}\ \bibnamefont {Kee}},\ and\ \bibinfo {author}
  {\bibfnamefont {Y.-J.}\ \bibnamefont {Kim}},\ }\bibfield  {title} {\bibinfo
  {title} {{$\ensuremath{\alpha}\ensuremath{-}{\mathrm{RuCl}}_{3}$: A
  spin-orbit assisted Mott insulator on a honeycomb lattice}},\ }\href
  {https://doi.org/10.1103/PhysRevB.90.041112} {\bibfield  {journal} {\bibinfo
  {journal} {Phys. Rev. B}\ }\textbf {\bibinfo {volume} {90}},\ \bibinfo
  {pages} {041112} (\bibinfo {year} {2014})}\BibitemShut {NoStop}%
\bibitem [{\citenamefont {Liu}\ and\ \citenamefont
  {Khaliullin}(2018)}]{LiuPRB2018}%
  \BibitemOpen
  \bibfield  {author} {\bibinfo {author} {\bibfnamefont {H.}~\bibnamefont
  {Liu}}\ and\ \bibinfo {author} {\bibfnamefont {G.}~\bibnamefont
  {Khaliullin}},\ }\bibfield  {title} {\bibinfo {title} {{Pseudospin exchange
  interactions in ${d}^{7}$ cobalt compounds: Possible realization of the
  Kitaev model}},\ }\href {https://doi.org/10.1103/PhysRevB.97.014407}
  {\bibfield  {journal} {\bibinfo  {journal} {Phys. Rev. B}\ }\textbf {\bibinfo
  {volume} {97}},\ \bibinfo {pages} {014407} (\bibinfo {year}
  {2018})}\BibitemShut {NoStop}%
\bibitem [{\citenamefont {Sano}\ \emph {et~al.}(2018)\citenamefont {Sano},
  \citenamefont {Kato},\ and\ \citenamefont {Motome}}]{SanoPRB2018}%
  \BibitemOpen
  \bibfield  {author} {\bibinfo {author} {\bibfnamefont {R.}~\bibnamefont
  {Sano}}, \bibinfo {author} {\bibfnamefont {Y.}~\bibnamefont {Kato}},\ and\
  \bibinfo {author} {\bibfnamefont {Y.}~\bibnamefont {Motome}},\ }\bibfield
  {title} {\bibinfo {title} {{Kitaev-Heisenberg Hamiltonian for high-spin
  ${d}^{7}$ Mott insulators}},\ }\href
  {https://doi.org/10.1103/PhysRevB.97.014408} {\bibfield  {journal} {\bibinfo
  {journal} {Phys. Rev. B}\ }\textbf {\bibinfo {volume} {97}},\ \bibinfo
  {pages} {014408} (\bibinfo {year} {2018})}\BibitemShut {NoStop}%
\bibitem [{\citenamefont {Motome}\ \emph {et~al.}(2020)\citenamefont {Motome},
  \citenamefont {Sano}, \citenamefont {Jang}, \citenamefont {Sugita},\ and\
  \citenamefont {Kato}}]{MotomeJPCM2020}%
  \BibitemOpen
  \bibfield  {author} {\bibinfo {author} {\bibfnamefont {Y.}~\bibnamefont
  {Motome}}, \bibinfo {author} {\bibfnamefont {R.}~\bibnamefont {Sano}},
  \bibinfo {author} {\bibfnamefont {S.}~\bibnamefont {Jang}}, \bibinfo {author}
  {\bibfnamefont {Y.}~\bibnamefont {Sugita}},\ and\ \bibinfo {author}
  {\bibfnamefont {Y.}~\bibnamefont {Kato}},\ }\bibfield  {title} {\bibinfo
  {title} {{Materials design of Kitaev spin liquids beyond the
  Jackeli{\textendash}Khaliullin mechanism}},\ }\href
  {https://doi.org/10.1088/1361-648x/ab8525} {\bibfield  {journal} {\bibinfo
  {journal} {Journal of Physics: Condensed Matter}\ }\textbf {\bibinfo {volume}
  {32}},\ \bibinfo {pages} {404001} (\bibinfo {year} {2020})}\BibitemShut
  {NoStop}%
\bibitem [{\citenamefont {Liu}\ \emph {et~al.}(2020)\citenamefont {Liu},
  \citenamefont {Chaloupka},\ and\ \citenamefont {Khaliullin}}]{LiuPRL2020}%
  \BibitemOpen
  \bibfield  {author} {\bibinfo {author} {\bibfnamefont {H.}~\bibnamefont
  {Liu}}, \bibinfo {author} {\bibfnamefont {J.}~\bibnamefont {Chaloupka}},\
  and\ \bibinfo {author} {\bibfnamefont {G.}~\bibnamefont {Khaliullin}},\
  }\bibfield  {title} {\bibinfo {title} {{Kitaev Spin Liquid in $3d$ Transition
  Metal Compounds}},\ }\href {https://doi.org/10.1103/PhysRevLett.125.047201}
  {\bibfield  {journal} {\bibinfo  {journal} {Phys. Rev. Lett.}\ }\textbf
  {\bibinfo {volume} {125}},\ \bibinfo {pages} {047201} (\bibinfo {year}
  {2020})}\BibitemShut {NoStop}%
\bibitem [{\citenamefont {Yan}\ \emph {et~al.}(2019)\citenamefont {Yan},
  \citenamefont {Okamoto}, \citenamefont {Wu}, \citenamefont {Zheng},
  \citenamefont {Zhou}, \citenamefont {Cao},\ and\ \citenamefont
  {McGuire}}]{YanPRM2019}%
  \BibitemOpen
  \bibfield  {author} {\bibinfo {author} {\bibfnamefont {J.-Q.}\ \bibnamefont
  {Yan}}, \bibinfo {author} {\bibfnamefont {S.}~\bibnamefont {Okamoto}},
  \bibinfo {author} {\bibfnamefont {Y.}~\bibnamefont {Wu}}, \bibinfo {author}
  {\bibfnamefont {Q.}~\bibnamefont {Zheng}}, \bibinfo {author} {\bibfnamefont
  {H.~D.}\ \bibnamefont {Zhou}}, \bibinfo {author} {\bibfnamefont {H.~B.}\
  \bibnamefont {Cao}},\ and\ \bibinfo {author} {\bibfnamefont {M.~A.}\
  \bibnamefont {McGuire}},\ }\bibfield  {title} {\bibinfo {title} {{Magnetic
  order in single crystals of
  ${\mathrm{Na}}_{3}{\mathrm{Co}}_{2}{\mathrm{SbO}}_{6}$ with a honeycomb
  arrangement of
  ${3\mathrm{d}}^{7}\phantom{\rule{0.28em}{0ex}}{\mathrm{Co}}^{2+}$ ions}},\
  }\href {https://doi.org/10.1103/PhysRevMaterials.3.074405} {\bibfield
  {journal} {\bibinfo  {journal} {Phys. Rev. Materials}\ }\textbf {\bibinfo
  {volume} {3}},\ \bibinfo {pages} {074405} (\bibinfo {year}
  {2019})}\BibitemShut {NoStop}%
\bibitem [{\citenamefont {Yao}\ and\ \citenamefont {Li}(2020)}]{YaoPRB2020}%
  \BibitemOpen
  \bibfield  {author} {\bibinfo {author} {\bibfnamefont {W.}~\bibnamefont
  {Yao}}\ and\ \bibinfo {author} {\bibfnamefont {Y.}~\bibnamefont {Li}},\
  }\bibfield  {title} {\bibinfo {title} {{Ferrimagnetism and anisotropic phase
  tunability by magnetic fields in
  ${\mathrm{Na}}_{2}{\mathrm{Co}}_{2}{\mathrm{TeO}}_{6}$}},\ }\href
  {https://doi.org/10.1103/PhysRevB.101.085120} {\bibfield  {journal} {\bibinfo
   {journal} {Phys. Rev. B}\ }\textbf {\bibinfo {volume} {101}},\ \bibinfo
  {pages} {085120} (\bibinfo {year} {2020})}\BibitemShut {NoStop}%
\bibitem [{\citenamefont {Zhong}\ \emph {et~al.}(2020)\citenamefont {Zhong},
  \citenamefont {Gao}, \citenamefont {Ong},\ and\ \citenamefont
  {Cava}}]{ZhongSA2020}%
  \BibitemOpen
  \bibfield  {author} {\bibinfo {author} {\bibfnamefont {R.}~\bibnamefont
  {Zhong}}, \bibinfo {author} {\bibfnamefont {T.}~\bibnamefont {Gao}}, \bibinfo
  {author} {\bibfnamefont {N.~P.}\ \bibnamefont {Ong}},\ and\ \bibinfo {author}
  {\bibfnamefont {R.~J.}\ \bibnamefont {Cava}},\ }\bibfield  {title} {\bibinfo
  {title} {{Weak-field induced nonmagnetic state in a Co-based honeycomb}},\
  }\href {https://doi.org/10.1126/sciadv.aay6953} {\bibfield  {journal}
  {\bibinfo  {journal} {Science Advances}\ }\textbf {\bibinfo {volume} {6}},\
  \bibinfo {pages} {eaay6953} (\bibinfo {year} {2020})}\BibitemShut {NoStop}%
\bibitem [{\citenamefont {Songvilay}\ \emph {et~al.}(2020)\citenamefont
  {Songvilay}, \citenamefont {Robert}, \citenamefont {Petit}, \citenamefont
  {Rodriguez-Rivera}, \citenamefont {Ratcliff}, \citenamefont {Damay},
  \citenamefont {Bal\'edent}, \citenamefont {Jim\'enez-Ruiz}, \citenamefont
  {Lejay}, \citenamefont {Pachoud}, \citenamefont {Hadj-Azzem}, \citenamefont
  {Simonet},\ and\ \citenamefont {Stock}}]{SongvilayPRB2020}%
  \BibitemOpen
  \bibfield  {author} {\bibinfo {author} {\bibfnamefont {M.}~\bibnamefont
  {Songvilay}}, \bibinfo {author} {\bibfnamefont {J.}~\bibnamefont {Robert}},
  \bibinfo {author} {\bibfnamefont {S.}~\bibnamefont {Petit}}, \bibinfo
  {author} {\bibfnamefont {J.~A.}\ \bibnamefont {Rodriguez-Rivera}}, \bibinfo
  {author} {\bibfnamefont {W.~D.}\ \bibnamefont {Ratcliff}}, \bibinfo {author}
  {\bibfnamefont {F.}~\bibnamefont {Damay}}, \bibinfo {author} {\bibfnamefont
  {V.}~\bibnamefont {Bal\'edent}}, \bibinfo {author} {\bibfnamefont
  {M.}~\bibnamefont {Jim\'enez-Ruiz}}, \bibinfo {author} {\bibfnamefont
  {P.}~\bibnamefont {Lejay}}, \bibinfo {author} {\bibfnamefont
  {E.}~\bibnamefont {Pachoud}}, \bibinfo {author} {\bibfnamefont
  {A.}~\bibnamefont {Hadj-Azzem}}, \bibinfo {author} {\bibfnamefont
  {V.}~\bibnamefont {Simonet}},\ and\ \bibinfo {author} {\bibfnamefont
  {C.}~\bibnamefont {Stock}},\ }\bibfield  {title} {\bibinfo {title} {{Kitaev
  interactions in the Co honeycomb antiferromagnets
  ${\mathrm{Na}}_{3}{\mathrm{Co}}_{2}{\mathrm{SbO}}_{6}$ and
  ${\mathrm{Na}}_{2}{\mathrm{Co}}_{2}{\mathrm{TeO}}_{6}$}},\ }\href
  {https://doi.org/10.1103/PhysRevB.102.224429} {\bibfield  {journal} {\bibinfo
   {journal} {Phys. Rev. B}\ }\textbf {\bibinfo {volume} {102}},\ \bibinfo
  {pages} {224429} (\bibinfo {year} {2020})}\BibitemShut {NoStop}%
\bibitem [{\citenamefont {Lin}\ \emph {et~al.}(2021)\citenamefont {Lin},
  \citenamefont {Jeong}, \citenamefont {Kim}, \citenamefont {Wang},
  \citenamefont {Huang}, \citenamefont {Masuda}, \citenamefont {Asai},
  \citenamefont {Itoh}, \citenamefont {G{\"u}nther}, \citenamefont {Russina},
  \citenamefont {Lu}, \citenamefont {Sheng}, \citenamefont {Wang},
  \citenamefont {Wang}, \citenamefont {Wang}, \citenamefont {Ren},
  \citenamefont {Xi}, \citenamefont {Tong}, \citenamefont {Ling}, \citenamefont
  {Liu}, \citenamefont {Wu}, \citenamefont {Mei}, \citenamefont {Qu},
  \citenamefont {Zhou}, \citenamefont {Park},\ and\ \citenamefont
  {Ma}}]{LinNC2021}%
  \BibitemOpen
  \bibfield  {author} {\bibinfo {author} {\bibfnamefont {G.}~\bibnamefont
  {Lin}}, \bibinfo {author} {\bibfnamefont {J.}~\bibnamefont {Jeong}}, \bibinfo
  {author} {\bibfnamefont {C.}~\bibnamefont {Kim}}, \bibinfo {author}
  {\bibfnamefont {Y.}~\bibnamefont {Wang}}, \bibinfo {author} {\bibfnamefont
  {Q.}~\bibnamefont {Huang}}, \bibinfo {author} {\bibfnamefont
  {T.}~\bibnamefont {Masuda}}, \bibinfo {author} {\bibfnamefont
  {S.}~\bibnamefont {Asai}}, \bibinfo {author} {\bibfnamefont {S.}~\bibnamefont
  {Itoh}}, \bibinfo {author} {\bibfnamefont {G.}~\bibnamefont {G{\"u}nther}},
  \bibinfo {author} {\bibfnamefont {M.}~\bibnamefont {Russina}}, \bibinfo
  {author} {\bibfnamefont {Z.}~\bibnamefont {Lu}}, \bibinfo {author}
  {\bibfnamefont {J.}~\bibnamefont {Sheng}}, \bibinfo {author} {\bibfnamefont
  {L.}~\bibnamefont {Wang}}, \bibinfo {author} {\bibfnamefont {J.}~\bibnamefont
  {Wang}}, \bibinfo {author} {\bibfnamefont {G.}~\bibnamefont {Wang}}, \bibinfo
  {author} {\bibfnamefont {Q.}~\bibnamefont {Ren}}, \bibinfo {author}
  {\bibfnamefont {C.}~\bibnamefont {Xi}}, \bibinfo {author} {\bibfnamefont
  {W.}~\bibnamefont {Tong}}, \bibinfo {author} {\bibfnamefont {L.}~\bibnamefont
  {Ling}}, \bibinfo {author} {\bibfnamefont {Z.}~\bibnamefont {Liu}}, \bibinfo
  {author} {\bibfnamefont {L.}~\bibnamefont {Wu}}, \bibinfo {author}
  {\bibfnamefont {J.}~\bibnamefont {Mei}}, \bibinfo {author} {\bibfnamefont
  {Z.}~\bibnamefont {Qu}}, \bibinfo {author} {\bibfnamefont {H.}~\bibnamefont
  {Zhou}}, \bibinfo {author} {\bibfnamefont {J.-G.}\ \bibnamefont {Park}},\
  and\ \bibinfo {author} {\bibfnamefont {J.}~\bibnamefont {Ma}},\ }\bibfield
  {title} {\bibinfo {title} {{Field-induced quantum spin disordered state in
  spin-1/2 honeycomb magnet Na$_2$Co$_2$TeO$_6$}},\ }\href
  {https://doi.org/10.1038/s41467-021-25567-7} {\bibfield  {journal} {\bibinfo
  {journal} {Nat. Commun.}\ }\textbf {\bibinfo {volume} {12}},\ \bibinfo
  {pages} {5559} (\bibinfo {year} {2021})}\BibitemShut {NoStop}%
\bibitem [{\citenamefont {Kim}\ \emph {et~al.}(2022{\natexlab{a}})\citenamefont
  {Kim}, \citenamefont {Jeong}, \citenamefont {Lin}, \citenamefont {Park},
  \citenamefont {Masuda}, \citenamefont {Asai}, \citenamefont {Itoh},
  \citenamefont {Kim}, \citenamefont {Zhou}, \citenamefont {Ma},\ and\
  \citenamefont {Park}}]{KimJPCM2022b}%
  \BibitemOpen
  \bibfield  {author} {\bibinfo {author} {\bibfnamefont {C.}~\bibnamefont
  {Kim}}, \bibinfo {author} {\bibfnamefont {J.}~\bibnamefont {Jeong}}, \bibinfo
  {author} {\bibfnamefont {G.}~\bibnamefont {Lin}}, \bibinfo {author}
  {\bibfnamefont {P.}~\bibnamefont {Park}}, \bibinfo {author} {\bibfnamefont
  {T.}~\bibnamefont {Masuda}}, \bibinfo {author} {\bibfnamefont
  {S.}~\bibnamefont {Asai}}, \bibinfo {author} {\bibfnamefont {S.}~\bibnamefont
  {Itoh}}, \bibinfo {author} {\bibfnamefont {H.-S.}\ \bibnamefont {Kim}},
  \bibinfo {author} {\bibfnamefont {H.}~\bibnamefont {Zhou}}, \bibinfo {author}
  {\bibfnamefont {J.}~\bibnamefont {Ma}},\ and\ \bibinfo {author}
  {\bibfnamefont {J.-G.}\ \bibnamefont {Park}},\ }\bibfield  {title} {\bibinfo
  {title} {{Antiferromagnetic Kitaev interaction Jeff= 1/2 cobalt honeycomb
  materials Na$_3$Co$_2$SbO$_6$ and Na$_2$Co$_2$TeO$_6$}},\ }\href
  {https://doi.org/10.1088/1361-648x/ac2644} {\bibfield  {journal} {\bibinfo
  {journal} {Journal of Physics: Condensed Matter}\ }\textbf {\bibinfo {volume}
  {34}},\ \bibinfo {pages} {045802} (\bibinfo {year}
  {2022}{\natexlab{a}})}\BibitemShut {NoStop}%
\bibitem [{\citenamefont {Kim}\ \emph {et~al.}(2022{\natexlab{b}})\citenamefont
  {Kim}, \citenamefont {Kim},\ and\ \citenamefont {Park}}]{KimJPCM2022}%
  \BibitemOpen
  \bibfield  {author} {\bibinfo {author} {\bibfnamefont {C.}~\bibnamefont
  {Kim}}, \bibinfo {author} {\bibfnamefont {H.-S.}\ \bibnamefont {Kim}},\ and\
  \bibinfo {author} {\bibfnamefont {J.-G.}\ \bibnamefont {Park}},\ }\bibfield
  {title} {\bibinfo {title} {{Spin-orbital entangled state and realization of
  Kitaev physics in 3d cobalt compounds: a progress report}},\ }\href
  {https://doi.org/10.1088/1361-648x/ac2d5d} {\bibfield  {journal} {\bibinfo
  {journal} {Journal of Physics: Condensed Matter}\ }\textbf {\bibinfo {volume}
  {34}},\ \bibinfo {pages} {023001} (\bibinfo {year}
  {2022}{\natexlab{b}})}\BibitemShut {NoStop}%
\bibitem [{\citenamefont {Das}\ \emph {et~al.}(2021)\citenamefont {Das},
  \citenamefont {Voleti}, \citenamefont {Saha-Dasgupta},\ and\ \citenamefont
  {Paramekanti}}]{DasPRB2021}%
  \BibitemOpen
  \bibfield  {author} {\bibinfo {author} {\bibfnamefont {S.}~\bibnamefont
  {Das}}, \bibinfo {author} {\bibfnamefont {S.}~\bibnamefont {Voleti}},
  \bibinfo {author} {\bibfnamefont {T.}~\bibnamefont {Saha-Dasgupta}},\ and\
  \bibinfo {author} {\bibfnamefont {A.}~\bibnamefont {Paramekanti}},\
  }\bibfield  {title} {\bibinfo {title} {{XY magnetism, Kitaev exchange, and
  long-range frustration in the ${J}_{\mathrm{eff}}=\frac{1}{2}$ honeycomb
  cobaltates}},\ }\href {https://doi.org/10.1103/PhysRevB.104.134425}
  {\bibfield  {journal} {\bibinfo  {journal} {Phys. Rev. B}\ }\textbf {\bibinfo
  {volume} {104}},\ \bibinfo {pages} {134425} (\bibinfo {year}
  {2021})}\BibitemShut {NoStop}%
\bibitem [{\citenamefont {Maksimov}\ \emph {et~al.}(2022)\citenamefont
  {Maksimov}, \citenamefont {Ushakov}, \citenamefont {Pchelkina}, \citenamefont
  {Li}, \citenamefont {Winter},\ and\ \citenamefont
  {Streltsov}}]{MaksimovPRB2022}%
  \BibitemOpen
  \bibfield  {author} {\bibinfo {author} {\bibfnamefont {P.~A.}\ \bibnamefont
  {Maksimov}}, \bibinfo {author} {\bibfnamefont {A.~V.}\ \bibnamefont
  {Ushakov}}, \bibinfo {author} {\bibfnamefont {Z.~V.}\ \bibnamefont
  {Pchelkina}}, \bibinfo {author} {\bibfnamefont {Y.}~\bibnamefont {Li}},
  \bibinfo {author} {\bibfnamefont {S.~M.}\ \bibnamefont {Winter}},\ and\
  \bibinfo {author} {\bibfnamefont {S.~V.}\ \bibnamefont {Streltsov}},\
  }\bibfield  {title} {\bibinfo {title} {{Ab initio guided minimal model for
  the ``Kitaev'' material ${\mathrm{BaCo}}_{2}$(${\mathrm{AsO}}_{4}{)}_{2}$:
  Importance of direct hopping, third-neighbor exchange, and quantum
  fluctuations}},\ }\href {https://doi.org/10.1103/PhysRevB.106.165131}
  {\bibfield  {journal} {\bibinfo  {journal} {Phys. Rev. B}\ }\textbf {\bibinfo
  {volume} {106}},\ \bibinfo {pages} {165131} (\bibinfo {year}
  {2022})}\BibitemShut {NoStop}%
\bibitem [{\citenamefont {Winter}(2022)}]{WinterJOP2022}%
  \BibitemOpen
  \bibfield  {author} {\bibinfo {author} {\bibfnamefont {S.~M.}\ \bibnamefont
  {Winter}},\ }\bibfield  {title} {\bibinfo {title} {{Magnetic couplings in
  edge-sharing high-spin $d^7$ compounds}},\ }\href
  {https://doi.org/10.1088/2515-7639/ac94f8} {\bibfield  {journal} {\bibinfo
  {journal} {Journal of Physics: Materials}\ }\textbf {\bibinfo {volume} {5}},\
  \bibinfo {pages} {045003} (\bibinfo {year} {2022})}\BibitemShut {NoStop}%
\bibitem [{\citenamefont {Pandey}\ and\ \citenamefont
  {Feng}(2022)}]{PandeyPRB2022}%
  \BibitemOpen
  \bibfield  {author} {\bibinfo {author} {\bibfnamefont {S.~K.}\ \bibnamefont
  {Pandey}}\ and\ \bibinfo {author} {\bibfnamefont {J.}~\bibnamefont {Feng}},\
  }\bibfield  {title} {\bibinfo {title} {{Spin interaction and magnetism in
  cobaltate Kitaev candidate materials: An ab initio and model Hamiltonian
  approach}},\ }\href {https://doi.org/10.1103/PhysRevB.106.174411} {\bibfield
  {journal} {\bibinfo  {journal} {Phys. Rev. B}\ }\textbf {\bibinfo {volume}
  {106}},\ \bibinfo {pages} {174411} (\bibinfo {year} {2022})}\BibitemShut
  {NoStop}%
\bibitem [{\citenamefont {Liu}\ and\ \citenamefont {Kee}(2023)}]{LiuPRB2023}%
  \BibitemOpen
  \bibfield  {author} {\bibinfo {author} {\bibfnamefont {X.}~\bibnamefont
  {Liu}}\ and\ \bibinfo {author} {\bibfnamefont {H.-Y.}\ \bibnamefont {Kee}},\
  }\bibfield  {title} {\bibinfo {title} {{Non-Kitaev versus Kitaev honeycomb
  cobaltates}},\ }\href {https://doi.org/10.1103/PhysRevB.107.054420}
  {\bibfield  {journal} {\bibinfo  {journal} {Phys. Rev. B}\ }\textbf {\bibinfo
  {volume} {107}},\ \bibinfo {pages} {054420} (\bibinfo {year}
  {2023})}\BibitemShut {NoStop}%
\bibitem [{\citenamefont {Halloran}\ \emph {et~al.}(2023)\citenamefont
  {Halloran}, \citenamefont {Desrochers}, \citenamefont {Zhang}, \citenamefont
  {Chen}, \citenamefont {Chern}, \citenamefont {Xu}, \citenamefont {Winn},
  \citenamefont {Graves-Brook}, \citenamefont {Stone}, \citenamefont
  {Kolesnikov}, \citenamefont {Qiu}, \citenamefont {Zhong}, \citenamefont
  {Cava}, \citenamefont {Kim},\ and\ \citenamefont
  {Broholm}}]{HalloranPNAS2023}%
  \BibitemOpen
  \bibfield  {author} {\bibinfo {author} {\bibfnamefont {T.}~\bibnamefont
  {Halloran}}, \bibinfo {author} {\bibfnamefont {F.}~\bibnamefont
  {Desrochers}}, \bibinfo {author} {\bibfnamefont {E.~Z.}\ \bibnamefont
  {Zhang}}, \bibinfo {author} {\bibfnamefont {T.}~\bibnamefont {Chen}},
  \bibinfo {author} {\bibfnamefont {L.~E.}\ \bibnamefont {Chern}}, \bibinfo
  {author} {\bibfnamefont {Z.}~\bibnamefont {Xu}}, \bibinfo {author}
  {\bibfnamefont {B.}~\bibnamefont {Winn}}, \bibinfo {author} {\bibfnamefont
  {M.}~\bibnamefont {Graves-Brook}}, \bibinfo {author} {\bibfnamefont {M.~B.}\
  \bibnamefont {Stone}}, \bibinfo {author} {\bibfnamefont {A.~I.}\ \bibnamefont
  {Kolesnikov}}, \bibinfo {author} {\bibfnamefont {Y.}~\bibnamefont {Qiu}},
  \bibinfo {author} {\bibfnamefont {R.}~\bibnamefont {Zhong}}, \bibinfo
  {author} {\bibfnamefont {R.}~\bibnamefont {Cava}}, \bibinfo {author}
  {\bibfnamefont {Y.~B.}\ \bibnamefont {Kim}},\ and\ \bibinfo {author}
  {\bibfnamefont {C.}~\bibnamefont {Broholm}},\ }\bibfield  {title} {\bibinfo
  {title} {{Geometrical frustration versus Kitaev interactions in
  BaCo$_2$(AsO$_4$)$_2$}},\ }\href {https://doi.org/10.1073/pnas.2215509119}
  {\bibfield  {journal} {\bibinfo  {journal} {Proceedings of the National
  Academy of Sciences}\ }\textbf {\bibinfo {volume} {120}},\ \bibinfo {pages}
  {e2215509119} (\bibinfo {year} {2023})}\BibitemShut {NoStop}%
\bibitem [{\citenamefont {Yao}\ \emph {et~al.}(2022)\citenamefont {Yao},
  \citenamefont {Iida}, \citenamefont {Kamazawa},\ and\ \citenamefont
  {Li}}]{YaoPRL2022}%
  \BibitemOpen
  \bibfield  {author} {\bibinfo {author} {\bibfnamefont {W.}~\bibnamefont
  {Yao}}, \bibinfo {author} {\bibfnamefont {K.}~\bibnamefont {Iida}}, \bibinfo
  {author} {\bibfnamefont {K.}~\bibnamefont {Kamazawa}},\ and\ \bibinfo
  {author} {\bibfnamefont {Y.}~\bibnamefont {Li}},\ }\bibfield  {title}
  {\bibinfo {title} {{Excitations in the Ordered and Paramagnetic States of
  Honeycomb Magnet ${\mathrm{Na}}_{2}{\mathrm{Co}}_{2}{\mathrm{TeO}}_{6}$}},\
  }\href {https://doi.org/10.1103/PhysRevLett.129.147202} {\bibfield  {journal}
  {\bibinfo  {journal} {Phys. Rev. Lett.}\ }\textbf {\bibinfo {volume} {129}},\
  \bibinfo {pages} {147202} (\bibinfo {year} {2022})}\BibitemShut {NoStop}%
\bibitem [{\citenamefont {Kr\"{u}ger}\ \emph {et~al.}(2022)\citenamefont
  {Kr\"{u}ger}, \citenamefont {Chen}, \citenamefont {Jin}, \citenamefont {Li},\
  and\ \citenamefont {Janssen}}]{KrugerArxiv2022}%
  \BibitemOpen
  \bibfield  {author} {\bibinfo {author} {\bibfnamefont {W.~G.~F.}\
  \bibnamefont {Kr\"{u}ger}}, \bibinfo {author} {\bibfnamefont
  {W.}~\bibnamefont {Chen}}, \bibinfo {author} {\bibfnamefont {X.}~\bibnamefont
  {Jin}}, \bibinfo {author} {\bibfnamefont {Y.}~\bibnamefont {Li}},\ and\
  \bibinfo {author} {\bibfnamefont {L.}~\bibnamefont {Janssen}},\ }\bibfield
  {title} {\bibinfo {title} {{Triple-Q order in Na$_2$Co$_2$TeO$_6$ from
  proximity to hidden-SU(2)-symmetric point}}\ }\href
  {https://doi.org/10.48550/arXiv.2211.16957} {10.48550/arXiv.2211.16957}
  (\bibinfo {year} {2022})\BibitemShut {NoStop}%
\bibitem [{\citenamefont {Wang}\ and\ \citenamefont
  {Liu}(2023)}]{WangArxiv2023}%
  \BibitemOpen
  \bibfield  {author} {\bibinfo {author} {\bibfnamefont {J.}~\bibnamefont
  {Wang}}\ and\ \bibinfo {author} {\bibfnamefont {Z.}~\bibnamefont {Liu}},\
  }\bibfield  {title} {\bibinfo {title} {{Effect of ring-exchange interactions
  in the extended Kitaev honeycomb model}}\ }\href
  {https://doi.org/10.48550/arXiv.2305.03258} {10.48550/arXiv.2305.03258}
  (\bibinfo {year} {2023})\BibitemShut {NoStop}%
\bibitem [{\citenamefont {Vavilova}\ \emph {et~al.}(2023)\citenamefont
  {Vavilova}, \citenamefont {Vasilchikova}, \citenamefont {Vasiliev},
  \citenamefont {Mikhailova}, \citenamefont {Nalbandyan}, \citenamefont
  {Zvereva},\ and\ \citenamefont {Streltsov}}]{VavilovaPRB2023}%
  \BibitemOpen
  \bibfield  {author} {\bibinfo {author} {\bibfnamefont {E.}~\bibnamefont
  {Vavilova}}, \bibinfo {author} {\bibfnamefont {T.}~\bibnamefont
  {Vasilchikova}}, \bibinfo {author} {\bibfnamefont {A.}~\bibnamefont
  {Vasiliev}}, \bibinfo {author} {\bibfnamefont {D.}~\bibnamefont
  {Mikhailova}}, \bibinfo {author} {\bibfnamefont {V.}~\bibnamefont
  {Nalbandyan}}, \bibinfo {author} {\bibfnamefont {E.}~\bibnamefont
  {Zvereva}},\ and\ \bibinfo {author} {\bibfnamefont {S.~V.}\ \bibnamefont
  {Streltsov}},\ }\bibfield  {title} {\bibinfo {title} {{Magnetic phase diagram
  and possible Kitaev-like behavior of the honeycomb-lattice antimonate
  ${\mathrm{Na}}_{3}{\mathrm{Co}}_{2}{\mathrm{SbO}}_{6}$}},\ }\href
  {https://doi.org/10.1103/PhysRevB.107.054411} {\bibfield  {journal} {\bibinfo
   {journal} {Phys. Rev. B}\ }\textbf {\bibinfo {volume} {107}},\ \bibinfo
  {pages} {054411} (\bibinfo {year} {2023})}\BibitemShut {NoStop}%
\bibitem [{\citenamefont {Kang}\ \emph {et~al.}(2023)\citenamefont {Kang},
  \citenamefont {Park}, \citenamefont {Song}, \citenamefont {Noh},
  \citenamefont {Choe}, \citenamefont {Kong}, \citenamefont {Kim},
  \citenamefont {Seo}, \citenamefont {Ko}, \citenamefont {Yi}, \citenamefont
  {Yoo}, \citenamefont {Park}, \citenamefont {Ok},\ and\ \citenamefont
  {Sohn}}]{KangPRB2023}%
  \BibitemOpen
  \bibfield  {author} {\bibinfo {author} {\bibfnamefont {B.}~\bibnamefont
  {Kang}}, \bibinfo {author} {\bibfnamefont {M.}~\bibnamefont {Park}}, \bibinfo
  {author} {\bibfnamefont {S.}~\bibnamefont {Song}}, \bibinfo {author}
  {\bibfnamefont {S.}~\bibnamefont {Noh}}, \bibinfo {author} {\bibfnamefont
  {D.}~\bibnamefont {Choe}}, \bibinfo {author} {\bibfnamefont {M.}~\bibnamefont
  {Kong}}, \bibinfo {author} {\bibfnamefont {M.}~\bibnamefont {Kim}}, \bibinfo
  {author} {\bibfnamefont {C.}~\bibnamefont {Seo}}, \bibinfo {author}
  {\bibfnamefont {E.~K.}\ \bibnamefont {Ko}}, \bibinfo {author} {\bibfnamefont
  {G.}~\bibnamefont {Yi}}, \bibinfo {author} {\bibfnamefont {J.-W.}\
  \bibnamefont {Yoo}}, \bibinfo {author} {\bibfnamefont {S.}~\bibnamefont
  {Park}}, \bibinfo {author} {\bibfnamefont {J.~M.}\ \bibnamefont {Ok}},\ and\
  \bibinfo {author} {\bibfnamefont {C.}~\bibnamefont {Sohn}},\ }\bibfield
  {title} {\bibinfo {title} {{Honeycomb oxide heterostructure as a candidate
  host for a Kitaev quantum spin liquid}},\ }\href
  {https://doi.org/10.1103/PhysRevB.107.075103} {\bibfield  {journal} {\bibinfo
   {journal} {Phys. Rev. B}\ }\textbf {\bibinfo {volume} {107}},\ \bibinfo
  {pages} {075103} (\bibinfo {year} {2023})}\BibitemShut {NoStop}%
\bibitem [{\citenamefont {Gu}\ \emph {et~al.}(2023)\citenamefont {Gu},
  \citenamefont {Li}, \citenamefont {Chen}, \citenamefont {Iida}, \citenamefont
  {Nakao}, \citenamefont {Munakata}, \citenamefont {Garlea}, \citenamefont
  {Li}, \citenamefont {Deng}, \citenamefont {Zaliznyak}, \citenamefont
  {Tranquada},\ and\ \citenamefont {Li}}]{GuArxiv2023}%
  \BibitemOpen
  \bibfield  {author} {\bibinfo {author} {\bibfnamefont {Y.}~\bibnamefont
  {Gu}}, \bibinfo {author} {\bibfnamefont {X.}~\bibnamefont {Li}}, \bibinfo
  {author} {\bibfnamefont {Y.}~\bibnamefont {Chen}}, \bibinfo {author}
  {\bibfnamefont {K.}~\bibnamefont {Iida}}, \bibinfo {author} {\bibfnamefont
  {A.}~\bibnamefont {Nakao}}, \bibinfo {author} {\bibfnamefont
  {K.}~\bibnamefont {Munakata}}, \bibinfo {author} {\bibfnamefont {V.~O.}\
  \bibnamefont {Garlea}}, \bibinfo {author} {\bibfnamefont {Y.}~\bibnamefont
  {Li}}, \bibinfo {author} {\bibfnamefont {G.}~\bibnamefont {Deng}}, \bibinfo
  {author} {\bibfnamefont {I.~A.}\ \bibnamefont {Zaliznyak}}, \bibinfo {author}
  {\bibfnamefont {J.~M.}\ \bibnamefont {Tranquada}},\ and\ \bibinfo {author}
  {\bibfnamefont {Y.}~\bibnamefont {Li}},\ }\bibfield  {title} {\bibinfo
  {title} {{Easy-plane multi-$\mathbf{q}$ magnetic ground state of
  Na$_3$Co$_2$CbO$_6$}}\ }\href {https://doi.org/10.48550/arXiv.2306.07175}
  {10.48550/arXiv.2306.07175} (\bibinfo {year} {2023})\BibitemShut {NoStop}%
\bibitem [{\citenamefont {Nair}\ \emph {et~al.}(2018)\citenamefont {Nair},
  \citenamefont {Brown}, \citenamefont {Coldren}, \citenamefont {Hester},
  \citenamefont {Gelfand}, \citenamefont {Podlesnyak}, \citenamefont {Huang},\
  and\ \citenamefont {Ross}}]{NairPRB2018}%
  \BibitemOpen
  \bibfield  {author} {\bibinfo {author} {\bibfnamefont {H.~S.}\ \bibnamefont
  {Nair}}, \bibinfo {author} {\bibfnamefont {J.~M.}\ \bibnamefont {Brown}},
  \bibinfo {author} {\bibfnamefont {E.}~\bibnamefont {Coldren}}, \bibinfo
  {author} {\bibfnamefont {G.}~\bibnamefont {Hester}}, \bibinfo {author}
  {\bibfnamefont {M.~P.}\ \bibnamefont {Gelfand}}, \bibinfo {author}
  {\bibfnamefont {A.}~\bibnamefont {Podlesnyak}}, \bibinfo {author}
  {\bibfnamefont {Q.}~\bibnamefont {Huang}},\ and\ \bibinfo {author}
  {\bibfnamefont {K.~A.}\ \bibnamefont {Ross}},\ }\bibfield  {title} {\bibinfo
  {title} {{Short-range order in the quantum XXZ honeycomb lattice material
  ${\mathrm{BaCo}}_{2}{({\mathrm{PO}}_{4})}_{2}$}},\ }\href
  {https://doi.org/10.1103/PhysRevB.97.134409} {\bibfield  {journal} {\bibinfo
  {journal} {Phys. Rev. B}\ }\textbf {\bibinfo {volume} {97}},\ \bibinfo
  {pages} {134409} (\bibinfo {year} {2018})}\BibitemShut {NoStop}%
\bibitem [{\citenamefont {Chen}\ \emph {et~al.}(2021)\citenamefont {Chen},
  \citenamefont {Li}, \citenamefont {Hu}, \citenamefont {Hu}, \citenamefont
  {Yue}, \citenamefont {Sutarto}, \citenamefont {He}, \citenamefont {Iida},
  \citenamefont {Kamazawa}, \citenamefont {Yu}, \citenamefont {Lin},\ and\
  \citenamefont {Li}}]{ChenPRB2021}%
  \BibitemOpen
  \bibfield  {author} {\bibinfo {author} {\bibfnamefont {W.}~\bibnamefont
  {Chen}}, \bibinfo {author} {\bibfnamefont {X.}~\bibnamefont {Li}}, \bibinfo
  {author} {\bibfnamefont {Z.}~\bibnamefont {Hu}}, \bibinfo {author}
  {\bibfnamefont {Z.}~\bibnamefont {Hu}}, \bibinfo {author} {\bibfnamefont
  {L.}~\bibnamefont {Yue}}, \bibinfo {author} {\bibfnamefont {R.}~\bibnamefont
  {Sutarto}}, \bibinfo {author} {\bibfnamefont {F.}~\bibnamefont {He}},
  \bibinfo {author} {\bibfnamefont {K.}~\bibnamefont {Iida}}, \bibinfo {author}
  {\bibfnamefont {K.}~\bibnamefont {Kamazawa}}, \bibinfo {author}
  {\bibfnamefont {W.}~\bibnamefont {Yu}}, \bibinfo {author} {\bibfnamefont
  {X.}~\bibnamefont {Lin}},\ and\ \bibinfo {author} {\bibfnamefont
  {Y.}~\bibnamefont {Li}},\ }\bibfield  {title} {\bibinfo {title} {{Spin-orbit
  phase behavior of ${\mathrm{Na}}_{2}{\mathrm{Co}}_{2}{\mathrm{TeO}}_{6}$ at
  low temperatures}},\ }\href {https://doi.org/10.1103/PhysRevB.103.L180404}
  {\bibfield  {journal} {\bibinfo  {journal} {Phys. Rev. B}\ }\textbf {\bibinfo
  {volume} {103}},\ \bibinfo {pages} {L180404} (\bibinfo {year}
  {2021})}\BibitemShut {NoStop}%
\bibitem [{\citenamefont {Li}\ \emph {et~al.}(2022)\citenamefont {Li},
  \citenamefont {Gu}, \citenamefont {Chen}, \citenamefont {Garlea},
  \citenamefont {Iida}, \citenamefont {Kamazawa}, \citenamefont {Li},
  \citenamefont {Deng}, \citenamefont {Xiao}, \citenamefont {Zheng},
  \citenamefont {Ye}, \citenamefont {Peng}, \citenamefont {Zaliznyak},
  \citenamefont {Tranquada},\ and\ \citenamefont {Li}}]{LiPRX2022}%
  \BibitemOpen
  \bibfield  {author} {\bibinfo {author} {\bibfnamefont {X.}~\bibnamefont
  {Li}}, \bibinfo {author} {\bibfnamefont {Y.}~\bibnamefont {Gu}}, \bibinfo
  {author} {\bibfnamefont {Y.}~\bibnamefont {Chen}}, \bibinfo {author}
  {\bibfnamefont {V.~O.}\ \bibnamefont {Garlea}}, \bibinfo {author}
  {\bibfnamefont {K.}~\bibnamefont {Iida}}, \bibinfo {author} {\bibfnamefont
  {K.}~\bibnamefont {Kamazawa}}, \bibinfo {author} {\bibfnamefont
  {Y.}~\bibnamefont {Li}}, \bibinfo {author} {\bibfnamefont {G.}~\bibnamefont
  {Deng}}, \bibinfo {author} {\bibfnamefont {Q.}~\bibnamefont {Xiao}}, \bibinfo
  {author} {\bibfnamefont {X.}~\bibnamefont {Zheng}}, \bibinfo {author}
  {\bibfnamefont {Z.}~\bibnamefont {Ye}}, \bibinfo {author} {\bibfnamefont
  {Y.}~\bibnamefont {Peng}}, \bibinfo {author} {\bibfnamefont {I.~A.}\
  \bibnamefont {Zaliznyak}}, \bibinfo {author} {\bibfnamefont {J.~M.}\
  \bibnamefont {Tranquada}},\ and\ \bibinfo {author} {\bibfnamefont
  {Y.}~\bibnamefont {Li}},\ }\bibfield  {title} {\bibinfo {title} {{Giant
  Magnetic In-Plane Anisotropy and Competing Instabilities in
  ${\mathrm{Na}}_{3}{\mathrm{Co}}_{2}{\mathrm{SbO}}_{6}$}},\ }\href
  {https://doi.org/10.1103/PhysRevX.12.041024} {\bibfield  {journal} {\bibinfo
  {journal} {Phys. Rev. X}\ }\textbf {\bibinfo {volume} {12}},\ \bibinfo
  {pages} {041024} (\bibinfo {year} {2022})}\BibitemShut {NoStop}%
\bibitem [{\citenamefont {Janssen}\ \emph {et~al.}(2017)\citenamefont
  {Janssen}, \citenamefont {Andrade},\ and\ \citenamefont
  {Vojta}}]{JanssenPRB2017}%
  \BibitemOpen
  \bibfield  {author} {\bibinfo {author} {\bibfnamefont {L.}~\bibnamefont
  {Janssen}}, \bibinfo {author} {\bibfnamefont {E.~C.}\ \bibnamefont
  {Andrade}},\ and\ \bibinfo {author} {\bibfnamefont {M.}~\bibnamefont
  {Vojta}},\ }\bibfield  {title} {\bibinfo {title} {{Magnetization processes of
  zigzag states on the honeycomb lattice: Identifying spin models for
  $\ensuremath{\alpha}\text{\ensuremath{-}}{\mathrm{RuCl}}_{3}$ and
  ${\mathrm{Na}}_{2}{\mathrm{IrO}}_{3}$}},\ }\href
  {https://doi.org/10.1103/PhysRevB.96.064430} {\bibfield  {journal} {\bibinfo
  {journal} {Phys. Rev. B}\ }\textbf {\bibinfo {volume} {96}},\ \bibinfo
  {pages} {064430} (\bibinfo {year} {2017})}\BibitemShut {NoStop}%
\bibitem [{SM()}]{SM}%
  \BibitemOpen
  \bibinfo {note} {See Supplemental Material at xxx for additional methods,
  data, and analyses.}\BibitemShut {Stop}%
\bibitem [{\citenamefont {Suter}\ and\ \citenamefont
  {Wojek}(2012)}]{SuterPP2012}%
  \BibitemOpen
  \bibfield  {author} {\bibinfo {author} {\bibfnamefont {A.}~\bibnamefont
  {Suter}}\ and\ \bibinfo {author} {\bibfnamefont {B.~M.}\ \bibnamefont
  {Wojek}},\ }\bibfield  {title} {\bibinfo {title} {{Musrfit: A Free
  Platform-Independent Framework for $\mu$SR Data Analysis}},\ }\href
  {https://doi.org/https://doi.org/10.1016/j.phpro.2012.04.042} {\bibfield
  {journal} {\bibinfo  {journal} {Physics Procedia}\ }\textbf {\bibinfo
  {volume} {30}},\ \bibinfo {pages} {69} (\bibinfo {year} {2012})}\BibitemShut
  {NoStop}%
\bibitem [{\citenamefont {Uemura}\ \emph {et~al.}(1985)\citenamefont {Uemura},
  \citenamefont {Yamazaki}, \citenamefont {Harshman}, \citenamefont {Senba},\
  and\ \citenamefont {Ansaldo}}]{UemuraPRB1985}%
  \BibitemOpen
  \bibfield  {author} {\bibinfo {author} {\bibfnamefont {Y.~J.}\ \bibnamefont
  {Uemura}}, \bibinfo {author} {\bibfnamefont {T.}~\bibnamefont {Yamazaki}},
  \bibinfo {author} {\bibfnamefont {D.~R.}\ \bibnamefont {Harshman}}, \bibinfo
  {author} {\bibfnamefont {M.}~\bibnamefont {Senba}},\ and\ \bibinfo {author}
  {\bibfnamefont {E.~J.}\ \bibnamefont {Ansaldo}},\ }\bibfield  {title}
  {\bibinfo {title} {{Muon-spin relaxation in AuFe and CuMn spin glasses}},\
  }\href {https://doi.org/10.1103/PhysRevB.31.546} {\bibfield  {journal}
  {\bibinfo  {journal} {Phys. Rev. B}\ }\textbf {\bibinfo {volume} {31}},\
  \bibinfo {pages} {546} (\bibinfo {year} {1985})}\BibitemShut {NoStop}%
\bibitem [{\citenamefont {Heffner}\ \emph {et~al.}(1996)\citenamefont
  {Heffner}, \citenamefont {Le}, \citenamefont {Hundley}, \citenamefont
  {Neumeier}, \citenamefont {Luke}, \citenamefont {Kojima}, \citenamefont
  {Nachumi}, \citenamefont {Uemura}, \citenamefont {MacLaughlin},\ and\
  \citenamefont {Cheong}}]{HeffnerPRL1996}%
  \BibitemOpen
  \bibfield  {author} {\bibinfo {author} {\bibfnamefont {R.~H.}\ \bibnamefont
  {Heffner}}, \bibinfo {author} {\bibfnamefont {L.~P.}\ \bibnamefont {Le}},
  \bibinfo {author} {\bibfnamefont {M.~F.}\ \bibnamefont {Hundley}}, \bibinfo
  {author} {\bibfnamefont {J.~J.}\ \bibnamefont {Neumeier}}, \bibinfo {author}
  {\bibfnamefont {G.~M.}\ \bibnamefont {Luke}}, \bibinfo {author}
  {\bibfnamefont {K.}~\bibnamefont {Kojima}}, \bibinfo {author} {\bibfnamefont
  {B.}~\bibnamefont {Nachumi}}, \bibinfo {author} {\bibfnamefont {Y.~J.}\
  \bibnamefont {Uemura}}, \bibinfo {author} {\bibfnamefont {D.~E.}\
  \bibnamefont {MacLaughlin}},\ and\ \bibinfo {author} {\bibfnamefont {S.~W.}\
  \bibnamefont {Cheong}},\ }\bibfield  {title} {\bibinfo {title}
  {{Ferromagnetic Ordering and Unusual Magnetic Ion Dynamics in
  ${\mathrm{La}}_{0.67}$${\mathrm{Ca}}_{0.33}$Mn${\mathrm{O}}_{3}$}},\ }\href
  {https://doi.org/10.1103/PhysRevLett.77.1869} {\bibfield  {journal} {\bibinfo
   {journal} {Phys. Rev. Lett.}\ }\textbf {\bibinfo {volume} {77}},\ \bibinfo
  {pages} {1869} (\bibinfo {year} {1996})}\BibitemShut {NoStop}%
\bibitem [{\citenamefont {Miao}\ \emph {et~al.}(2019)\citenamefont {Miao},
  \citenamefont {Wang}, \citenamefont {Zhu}, \citenamefont {Liu}, \citenamefont
  {Liu}, \citenamefont {Hu}, \citenamefont {Li}, \citenamefont {Tan},
  \citenamefont {Koda}, \citenamefont {Zhu}, \citenamefont {Feng},
  \citenamefont {Su}, \citenamefont {Kamiyama}, \citenamefont {Xiao},\ and\
  \citenamefont {Pan}}]{MiaoAPL2019}%
  \BibitemOpen
  \bibfield  {author} {\bibinfo {author} {\bibfnamefont {P.}~\bibnamefont
  {Miao}}, \bibinfo {author} {\bibfnamefont {R.}~\bibnamefont {Wang}}, \bibinfo
  {author} {\bibfnamefont {W.}~\bibnamefont {Zhu}}, \bibinfo {author}
  {\bibfnamefont {J.}~\bibnamefont {Liu}}, \bibinfo {author} {\bibfnamefont
  {T.}~\bibnamefont {Liu}}, \bibinfo {author} {\bibfnamefont {J.}~\bibnamefont
  {Hu}}, \bibinfo {author} {\bibfnamefont {S.}~\bibnamefont {Li}}, \bibinfo
  {author} {\bibfnamefont {Z.}~\bibnamefont {Tan}}, \bibinfo {author}
  {\bibfnamefont {A.}~\bibnamefont {Koda}}, \bibinfo {author} {\bibfnamefont
  {F.}~\bibnamefont {Zhu}}, \bibinfo {author} {\bibfnamefont {E.}~\bibnamefont
  {Feng}}, \bibinfo {author} {\bibfnamefont {Y.}~\bibnamefont {Su}}, \bibinfo
  {author} {\bibfnamefont {T.}~\bibnamefont {Kamiyama}}, \bibinfo {author}
  {\bibfnamefont {Y.}~\bibnamefont {Xiao}},\ and\ \bibinfo {author}
  {\bibfnamefont {F.}~\bibnamefont {Pan}},\ }\bibfield  {title} {\bibinfo
  {title} {{Revealing magnetic ground state of a layered cathode material by
  muon spin relaxation and neutron scattering experiments}},\ }\href
  {https://doi.org/10.1063/1.5096620} {\bibfield  {journal} {\bibinfo
  {journal} {Applied Physics Letters}\ }\textbf {\bibinfo {volume} {114}},\
  \bibinfo {pages} {203901} (\bibinfo {year} {2019})}\BibitemShut {NoStop}%
\bibitem [{\citenamefont {Campbell}\ \emph {et~al.}(1994)\citenamefont
  {Campbell}, \citenamefont {Amato}, \citenamefont {Gygax}, \citenamefont
  {Herlach}, \citenamefont {Schenck}, \citenamefont {Cywinski},\ and\
  \citenamefont {Kilcoyne}}]{CampbellPRL1994}%
  \BibitemOpen
  \bibfield  {author} {\bibinfo {author} {\bibfnamefont {I.~A.}\ \bibnamefont
  {Campbell}}, \bibinfo {author} {\bibfnamefont {A.}~\bibnamefont {Amato}},
  \bibinfo {author} {\bibfnamefont {F.~N.}\ \bibnamefont {Gygax}}, \bibinfo
  {author} {\bibfnamefont {D.}~\bibnamefont {Herlach}}, \bibinfo {author}
  {\bibfnamefont {A.}~\bibnamefont {Schenck}}, \bibinfo {author} {\bibfnamefont
  {R.}~\bibnamefont {Cywinski}},\ and\ \bibinfo {author} {\bibfnamefont
  {S.~H.}\ \bibnamefont {Kilcoyne}},\ }\bibfield  {title} {\bibinfo {title}
  {{Dynamics in canonical spin glasses observed by muon spin depolarization}},\
  }\href {https://doi.org/10.1103/PhysRevLett.72.1291} {\bibfield  {journal}
  {\bibinfo  {journal} {Phys. Rev. Lett.}\ }\textbf {\bibinfo {volume} {72}},\
  \bibinfo {pages} {1291} (\bibinfo {year} {1994})}\BibitemShut {NoStop}%
\bibitem [{\citenamefont {Keren}\ \emph {et~al.}(1996)\citenamefont {Keren},
  \citenamefont {Mendels}, \citenamefont {Campbell},\ and\ \citenamefont
  {Lord}}]{KerenPRL1996}%
  \BibitemOpen
  \bibfield  {author} {\bibinfo {author} {\bibfnamefont {A.}~\bibnamefont
  {Keren}}, \bibinfo {author} {\bibfnamefont {P.}~\bibnamefont {Mendels}},
  \bibinfo {author} {\bibfnamefont {I.~A.}\ \bibnamefont {Campbell}},\ and\
  \bibinfo {author} {\bibfnamefont {J.}~\bibnamefont {Lord}},\ }\bibfield
  {title} {\bibinfo {title} {{Probing the Spin-Spin Dynamical Autocorrelation
  Function in a Spin Glass above ${T}_{g}$ via Muon Spin Relaxation}},\ }\href
  {https://doi.org/10.1103/PhysRevLett.77.1386} {\bibfield  {journal} {\bibinfo
   {journal} {Phys. Rev. Lett.}\ }\textbf {\bibinfo {volume} {77}},\ \bibinfo
  {pages} {1386} (\bibinfo {year} {1996})}\BibitemShut {NoStop}%
\bibitem [{\citenamefont {Dunsiger}\ \emph {et~al.}(1996)\citenamefont
  {Dunsiger}, \citenamefont {Kiefl}, \citenamefont {Chow}, \citenamefont
  {Gaulin}, \citenamefont {Gingras}, \citenamefont {Greedan}, \citenamefont
  {Keren}, \citenamefont {Kojima}, \citenamefont {Luke}, \citenamefont
  {MacFarlane}, \citenamefont {Raju}, \citenamefont {Sonier}, \citenamefont
  {Uemura},\ and\ \citenamefont {Wu}}]{DunsigerJAP1996}%
  \BibitemOpen
  \bibfield  {author} {\bibinfo {author} {\bibfnamefont {S.~R.}\ \bibnamefont
  {Dunsiger}}, \bibinfo {author} {\bibfnamefont {R.~F.}\ \bibnamefont {Kiefl}},
  \bibinfo {author} {\bibfnamefont {K.~H.}\ \bibnamefont {Chow}}, \bibinfo
  {author} {\bibfnamefont {B.~D.}\ \bibnamefont {Gaulin}}, \bibinfo {author}
  {\bibfnamefont {M.~J.~P.}\ \bibnamefont {Gingras}}, \bibinfo {author}
  {\bibfnamefont {J.~E.}\ \bibnamefont {Greedan}}, \bibinfo {author}
  {\bibfnamefont {A.}~\bibnamefont {Keren}}, \bibinfo {author} {\bibfnamefont
  {K.}~\bibnamefont {Kojima}}, \bibinfo {author} {\bibfnamefont {G.~M.}\
  \bibnamefont {Luke}}, \bibinfo {author} {\bibfnamefont {W.~A.}\ \bibnamefont
  {MacFarlane}}, \bibinfo {author} {\bibfnamefont {N.~P.}\ \bibnamefont
  {Raju}}, \bibinfo {author} {\bibfnamefont {J.~E.}\ \bibnamefont {Sonier}},
  \bibinfo {author} {\bibfnamefont {Y.~J.}\ \bibnamefont {Uemura}},\ and\
  \bibinfo {author} {\bibfnamefont {W.~D.}\ \bibnamefont {Wu}},\ }\bibfield
  {title} {\bibinfo {title} {{Low temperature spin dynamics of geometrically
  frustrated antiferromagnets Y$_2$Mo$_2$O$_7$ and
  Y$_2$Mo$_{1.6}$Ti$_{0.4}$O$_7$ studied by muon spin relaxation}},\ }\href
  {https://doi.org/10.1063/1.361908} {\bibfield  {journal} {\bibinfo  {journal}
  {Journal of Applied Physics}\ }\textbf {\bibinfo {volume} {79}},\ \bibinfo
  {pages} {6636} (\bibinfo {year} {1996})}\BibitemShut {NoStop}%
\bibitem [{\citenamefont {Gardner}\ \emph {et~al.}(1999)\citenamefont
  {Gardner}, \citenamefont {Gaulin}, \citenamefont {Lee}, \citenamefont
  {Broholm}, \citenamefont {Raju},\ and\ \citenamefont
  {Greedan}}]{GardnerPRL1999}%
  \BibitemOpen
  \bibfield  {author} {\bibinfo {author} {\bibfnamefont {J.~S.}\ \bibnamefont
  {Gardner}}, \bibinfo {author} {\bibfnamefont {B.~D.}\ \bibnamefont {Gaulin}},
  \bibinfo {author} {\bibfnamefont {S.-H.}\ \bibnamefont {Lee}}, \bibinfo
  {author} {\bibfnamefont {C.}~\bibnamefont {Broholm}}, \bibinfo {author}
  {\bibfnamefont {N.~P.}\ \bibnamefont {Raju}},\ and\ \bibinfo {author}
  {\bibfnamefont {J.~E.}\ \bibnamefont {Greedan}},\ }\bibfield  {title}
  {\bibinfo {title} {{Glassy Statics and Dynamics in the Chemically Ordered
  Pyrochlore Antiferromagnet ${Y}_{2}{\mathrm{Mo}}_{2}{O}_{7}$}},\ }\href
  {https://doi.org/10.1103/PhysRevLett.83.211} {\bibfield  {journal} {\bibinfo
  {journal} {Phys. Rev. Lett.}\ }\textbf {\bibinfo {volume} {83}},\ \bibinfo
  {pages} {211} (\bibinfo {year} {1999})}\BibitemShut {NoStop}%
\bibitem [{\citenamefont {Dey}\ \emph {et~al.}(2023)\citenamefont {Dey},
  \citenamefont {Ishida}, \citenamefont {Okabe}, \citenamefont {Hiraishi},
  \citenamefont {Koda}, \citenamefont {Honda}, \citenamefont {Yamaura},
  \citenamefont {Kageyama},\ and\ \citenamefont {Kadono}}]{DeyPRB2023}%
  \BibitemOpen
  \bibfield  {author} {\bibinfo {author} {\bibfnamefont {S.~K.}\ \bibnamefont
  {Dey}}, \bibinfo {author} {\bibfnamefont {K.}~\bibnamefont {Ishida}},
  \bibinfo {author} {\bibfnamefont {H.}~\bibnamefont {Okabe}}, \bibinfo
  {author} {\bibfnamefont {M.}~\bibnamefont {Hiraishi}}, \bibinfo {author}
  {\bibfnamefont {A.}~\bibnamefont {Koda}}, \bibinfo {author} {\bibfnamefont
  {T.}~\bibnamefont {Honda}}, \bibinfo {author} {\bibfnamefont
  {J.}~\bibnamefont {Yamaura}}, \bibinfo {author} {\bibfnamefont
  {H.}~\bibnamefont {Kageyama}},\ and\ \bibinfo {author} {\bibfnamefont
  {R.}~\bibnamefont {Kadono}},\ }\bibfield  {title} {\bibinfo {title} {{Local
  spin dynamics in the geometrically frustrated Mo pyrochlore antiferromagnet
  ${\mathrm{Lu}}_{2}{\mathrm{Mo}}_{2}{\mathrm{O}}_{5\ensuremath{-}y}{\mathrm{N}}_{2}$}},\
  }\href {https://doi.org/10.1103/PhysRevB.107.024407} {\bibfield  {journal}
  {\bibinfo  {journal} {Phys. Rev. B}\ }\textbf {\bibinfo {volume} {107}},\
  \bibinfo {pages} {024407} (\bibinfo {year} {2023})}\BibitemShut {NoStop}%
\bibitem [{\citenamefont {Gubbens}\ \emph {et~al.}(1994)\citenamefont
  {Gubbens}, \citenamefont {Moolenaar}, \citenamefont {Dalmas De~R{\'e}otier},
  \citenamefont {Yaouanc}, \citenamefont {Kayzel}, \citenamefont {Franse},
  \citenamefont {Prokes}, \citenamefont {Snel}, \citenamefont {Bonville},
  \citenamefont {Hodges}, \citenamefont {Imbert},\ and\ \citenamefont
  {Pari}}]{Gubbens1994}%
  \BibitemOpen
  \bibfield  {author} {\bibinfo {author} {\bibfnamefont {P.~C.~M.}\
  \bibnamefont {Gubbens}}, \bibinfo {author} {\bibfnamefont {A.~A.}\
  \bibnamefont {Moolenaar}}, \bibinfo {author} {\bibfnamefont {P.}~\bibnamefont
  {Dalmas De~R{\'e}otier}}, \bibinfo {author} {\bibfnamefont {A.}~\bibnamefont
  {Yaouanc}}, \bibinfo {author} {\bibfnamefont {F.}~\bibnamefont {Kayzel}},
  \bibinfo {author} {\bibfnamefont {J.~J.~M.}\ \bibnamefont {Franse}}, \bibinfo
  {author} {\bibfnamefont {K.}~\bibnamefont {Prokes}}, \bibinfo {author}
  {\bibfnamefont {C.~E.}\ \bibnamefont {Snel}}, \bibinfo {author}
  {\bibfnamefont {P.}~\bibnamefont {Bonville}}, \bibinfo {author}
  {\bibfnamefont {J.~A.}\ \bibnamefont {Hodges}}, \bibinfo {author}
  {\bibfnamefont {P.}~\bibnamefont {Imbert}},\ and\ \bibinfo {author}
  {\bibfnamefont {P.}~\bibnamefont {Pari}},\ }\bibfield  {title} {\bibinfo
  {title} {{Spin dynamics in RENi$_5$ ferromagnets by $\mu$SR measurements}},\
  }\href {https://doi.org/10.1007/BF02069428} {\bibfield  {journal} {\bibinfo
  {journal} {Hyperfine Interactions}\ }\textbf {\bibinfo {volume} {85}},\
  \bibinfo {pages} {239} (\bibinfo {year} {1994})}\BibitemShut {NoStop}%
\bibitem [{\citenamefont {Miao}\ \emph {et~al.}(2021)\citenamefont {Miao},
  \citenamefont {Tan}, \citenamefont {Lee}, \citenamefont {Ishikawa},
  \citenamefont {Torii}, \citenamefont {Yonemura}, \citenamefont {Koda},
  \citenamefont {Komatsu}, \citenamefont {Machida}, \citenamefont
  {Sano-Furukawa}, \citenamefont {Hattori}, \citenamefont {Lin}, \citenamefont
  {Li}, \citenamefont {Mochiku}, \citenamefont {Kikuchi}, \citenamefont
  {Kawashima}, \citenamefont {Takahashi}, \citenamefont {Huang}, \citenamefont
  {Itoh}, \citenamefont {Kadono}, \citenamefont {Wang}, \citenamefont {Pan},
  \citenamefont {Yamauchi},\ and\ \citenamefont {Kamiyama}}]{MiaoPRB2021}%
  \BibitemOpen
  \bibfield  {author} {\bibinfo {author} {\bibfnamefont {P.}~\bibnamefont
  {Miao}}, \bibinfo {author} {\bibfnamefont {Z.}~\bibnamefont {Tan}}, \bibinfo
  {author} {\bibfnamefont {S.}~\bibnamefont {Lee}}, \bibinfo {author}
  {\bibfnamefont {Y.}~\bibnamefont {Ishikawa}}, \bibinfo {author}
  {\bibfnamefont {S.}~\bibnamefont {Torii}}, \bibinfo {author} {\bibfnamefont
  {M.}~\bibnamefont {Yonemura}}, \bibinfo {author} {\bibfnamefont
  {A.}~\bibnamefont {Koda}}, \bibinfo {author} {\bibfnamefont {K.}~\bibnamefont
  {Komatsu}}, \bibinfo {author} {\bibfnamefont {S.}~\bibnamefont {Machida}},
  \bibinfo {author} {\bibfnamefont {A.}~\bibnamefont {Sano-Furukawa}}, \bibinfo
  {author} {\bibfnamefont {T.}~\bibnamefont {Hattori}}, \bibinfo {author}
  {\bibfnamefont {X.}~\bibnamefont {Lin}}, \bibinfo {author} {\bibfnamefont
  {K.}~\bibnamefont {Li}}, \bibinfo {author} {\bibfnamefont {T.}~\bibnamefont
  {Mochiku}}, \bibinfo {author} {\bibfnamefont {R.}~\bibnamefont {Kikuchi}},
  \bibinfo {author} {\bibfnamefont {C.}~\bibnamefont {Kawashima}}, \bibinfo
  {author} {\bibfnamefont {H.}~\bibnamefont {Takahashi}}, \bibinfo {author}
  {\bibfnamefont {Q.}~\bibnamefont {Huang}}, \bibinfo {author} {\bibfnamefont
  {S.}~\bibnamefont {Itoh}}, \bibinfo {author} {\bibfnamefont {R.}~\bibnamefont
  {Kadono}}, \bibinfo {author} {\bibfnamefont {Y.}~\bibnamefont {Wang}},
  \bibinfo {author} {\bibfnamefont {F.}~\bibnamefont {Pan}}, \bibinfo {author}
  {\bibfnamefont {K.}~\bibnamefont {Yamauchi}},\ and\ \bibinfo {author}
  {\bibfnamefont {T.}~\bibnamefont {Kamiyama}},\ }\bibfield  {title} {\bibinfo
  {title} {Origin of magnetovolume effect in a cobaltite},\ }\href
  {https://doi.org/10.1103/PhysRevB.103.094302} {\bibfield  {journal} {\bibinfo
   {journal} {Phys. Rev. B}\ }\textbf {\bibinfo {volume} {103}},\ \bibinfo
  {pages} {094302} (\bibinfo {year} {2021})}\BibitemShut {NoStop}%
\bibitem [{\citenamefont {Yao}\ \emph {et~al.}(2023)\citenamefont {Yao},
  \citenamefont {Zhao}, \citenamefont {Qiu}, \citenamefont {Balz},
  \citenamefont {Stewart}, \citenamefont {Lynn},\ and\ \citenamefont
  {Li}}]{YaoPRR2023}%
  \BibitemOpen
  \bibfield  {author} {\bibinfo {author} {\bibfnamefont {W.}~\bibnamefont
  {Yao}}, \bibinfo {author} {\bibfnamefont {Y.}~\bibnamefont {Zhao}}, \bibinfo
  {author} {\bibfnamefont {Y.}~\bibnamefont {Qiu}}, \bibinfo {author}
  {\bibfnamefont {C.}~\bibnamefont {Balz}}, \bibinfo {author} {\bibfnamefont
  {J.~R.}\ \bibnamefont {Stewart}}, \bibinfo {author} {\bibfnamefont {J.~W.}\
  \bibnamefont {Lynn}},\ and\ \bibinfo {author} {\bibfnamefont
  {Y.}~\bibnamefont {Li}},\ }\bibfield  {title} {\bibinfo {title} {{Magnetic
  ground state of the Kitaev
  ${\mathrm{Na}}_{2}{\mathrm{Co}}_{2}{\mathrm{TeO}}_{6}$ spin liquid
  candidate}},\ }\href {https://doi.org/10.1103/PhysRevResearch.5.L022045}
  {\bibfield  {journal} {\bibinfo  {journal} {Phys. Rev. Res.}\ }\textbf
  {\bibinfo {volume} {5}},\ \bibinfo {pages} {L022045} (\bibinfo {year}
  {2023})}\BibitemShut {NoStop}%
\bibitem [{\citenamefont {Lefran\ifmmode~\mbox{\c{c}}\else \c{c}\fi{}ois}\
  \emph {et~al.}(2016)\citenamefont {Lefran\ifmmode~\mbox{\c{c}}\else
  \c{c}\fi{}ois}, \citenamefont {Songvilay}, \citenamefont {Robert},
  \citenamefont {Nataf}, \citenamefont {Jordan}, \citenamefont {Chaix},
  \citenamefont {Colin}, \citenamefont {Lejay}, \citenamefont {Hadj-Azzem},
  \citenamefont {Ballou},\ and\ \citenamefont {Simonet}}]{LefrancoisPRB2016}%
  \BibitemOpen
  \bibfield  {author} {\bibinfo {author} {\bibfnamefont {E.}~\bibnamefont
  {Lefran\ifmmode~\mbox{\c{c}}\else \c{c}\fi{}ois}}, \bibinfo {author}
  {\bibfnamefont {M.}~\bibnamefont {Songvilay}}, \bibinfo {author}
  {\bibfnamefont {J.}~\bibnamefont {Robert}}, \bibinfo {author} {\bibfnamefont
  {G.}~\bibnamefont {Nataf}}, \bibinfo {author} {\bibfnamefont
  {E.}~\bibnamefont {Jordan}}, \bibinfo {author} {\bibfnamefont
  {L.}~\bibnamefont {Chaix}}, \bibinfo {author} {\bibfnamefont {C.~V.}\
  \bibnamefont {Colin}}, \bibinfo {author} {\bibfnamefont {P.}~\bibnamefont
  {Lejay}}, \bibinfo {author} {\bibfnamefont {A.}~\bibnamefont {Hadj-Azzem}},
  \bibinfo {author} {\bibfnamefont {R.}~\bibnamefont {Ballou}},\ and\ \bibinfo
  {author} {\bibfnamefont {V.}~\bibnamefont {Simonet}},\ }\bibfield  {title}
  {\bibinfo {title} {{Magnetic properties of the honeycomb oxide
  ${\mathrm{Na}}_{2}{\mathrm{Co}}_{2}{\mathrm{TeO}}_{6}$}},\ }\href
  {https://doi.org/10.1103/PhysRevB.94.214416} {\bibfield  {journal} {\bibinfo
  {journal} {Phys. Rev. B}\ }\textbf {\bibinfo {volume} {94}},\ \bibinfo
  {pages} {214416} (\bibinfo {year} {2016})}\BibitemShut {NoStop}%
\bibitem [{\citenamefont {Bera}\ \emph {et~al.}(2017)\citenamefont {Bera},
  \citenamefont {Yusuf}, \citenamefont {Kumar},\ and\ \citenamefont
  {Ritter}}]{BeraPRB2017}%
  \BibitemOpen
  \bibfield  {author} {\bibinfo {author} {\bibfnamefont {A.~K.}\ \bibnamefont
  {Bera}}, \bibinfo {author} {\bibfnamefont {S.~M.}\ \bibnamefont {Yusuf}},
  \bibinfo {author} {\bibfnamefont {A.}~\bibnamefont {Kumar}},\ and\ \bibinfo
  {author} {\bibfnamefont {C.}~\bibnamefont {Ritter}},\ }\bibfield  {title}
  {\bibinfo {title} {{Zigzag antiferromagnetic ground state with anisotropic
  correlation lengths in the quasi-two-dimensional honeycomb lattice compound
  $\mathrm{N}{\mathrm{a}}_{2}\mathrm{C}{\mathrm{o}}_{2}\mathrm{Te}{\mathrm{O}}_{6}$}},\
  }\href {https://doi.org/10.1103/PhysRevB.95.094424} {\bibfield  {journal}
  {\bibinfo  {journal} {Phys. Rev. B}\ }\textbf {\bibinfo {volume} {95}},\
  \bibinfo {pages} {094424} (\bibinfo {year} {2017})}\BibitemShut {NoStop}%
\end{thebibliography}%

%%%%%%%%%% Merge with supplemental materials %%%%%%%%%%
\pagebreak
\pagebreak

\widetext
\begin{center}
\textbf{\large Supplemental Material for ``Persistent spin dynamics in magnetically ordered honeycomb cobalt oxides''}
\end{center}

\makeatletter
\renewcommand{\thefigure}{S\arabic{figure}}

\begin{figure}[!h]
\includegraphics[width=5in]{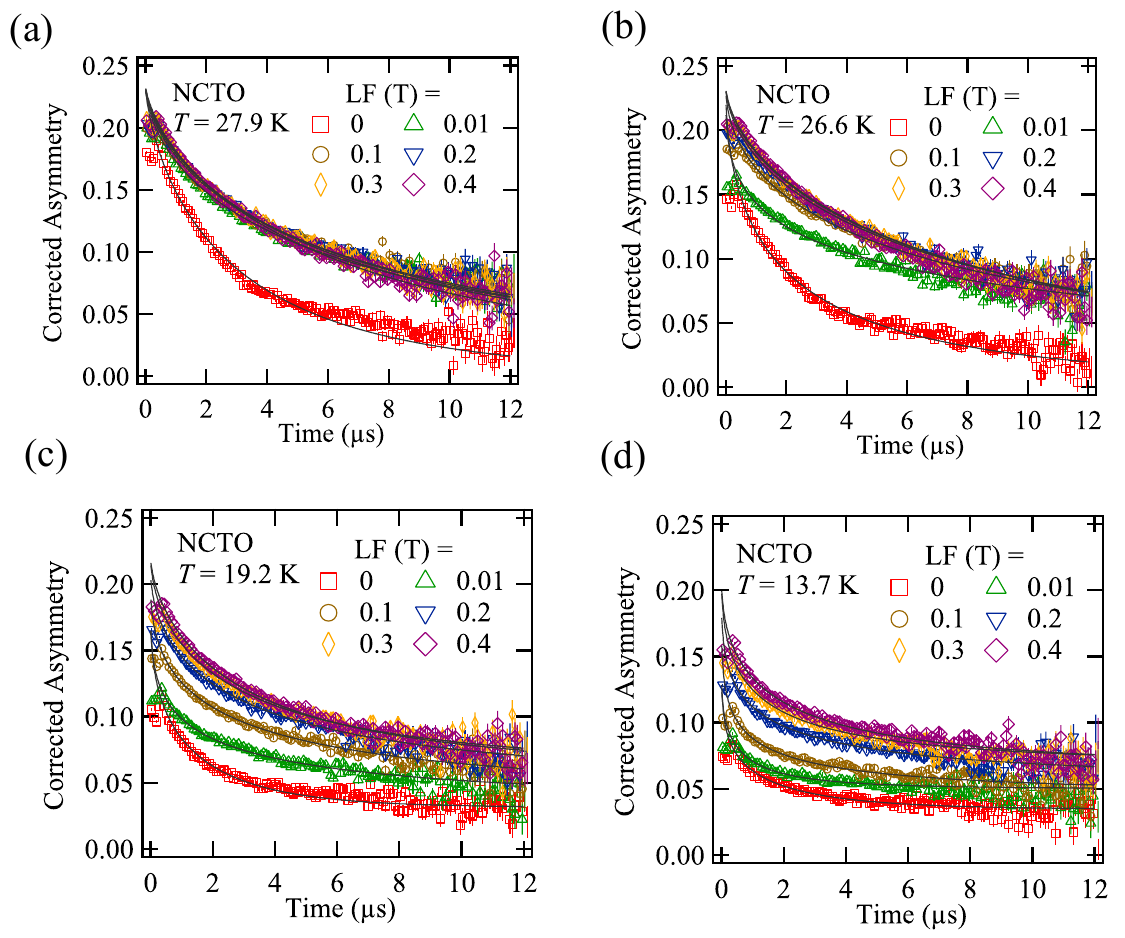}
\caption{$\mu$SR spectra of  \ch{Na_2Co_2TeO_6} (NCTO)  under various longitudinal fields (LFs) at  $T$ = 27.9 K (a), 26.6 K (b), 19.2 K (c) and 13.7 K (d).}
\label{figS1}
\end{figure}

\begin{figure}[!h]
\includegraphics[width=5in]{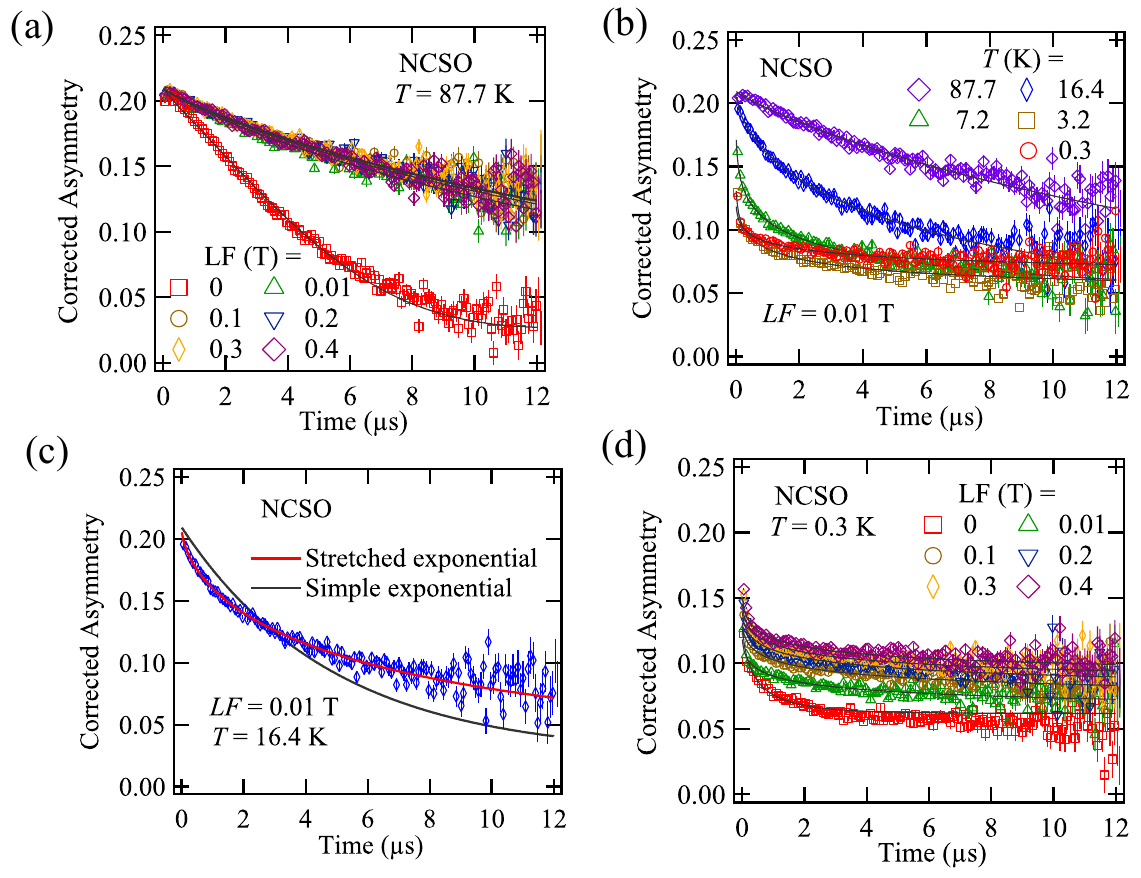}
\caption{(a) $\mu$SR spectra of \ch{Na_3Co_2SbO_6} (NCSO) measured far above $T_\mathrm{N}$. A longitudinal field (LF) of 0.01 T decouples the nuclear dipolar fields. (b) Near-zero-field $\mu$SR spectra measured at selected temperatures. (c) Demonstration of stretched exponential fitting of the data obtained in near-zero field and just above $T_\mathrm{N}$. (d) Lowest-temperature spectra under various LFs, revealing a dual static and dynamic nature of the relaxation.}
\label{figS2}
\end{figure}

\begin{figure}[!h]
\includegraphics[width=5in]{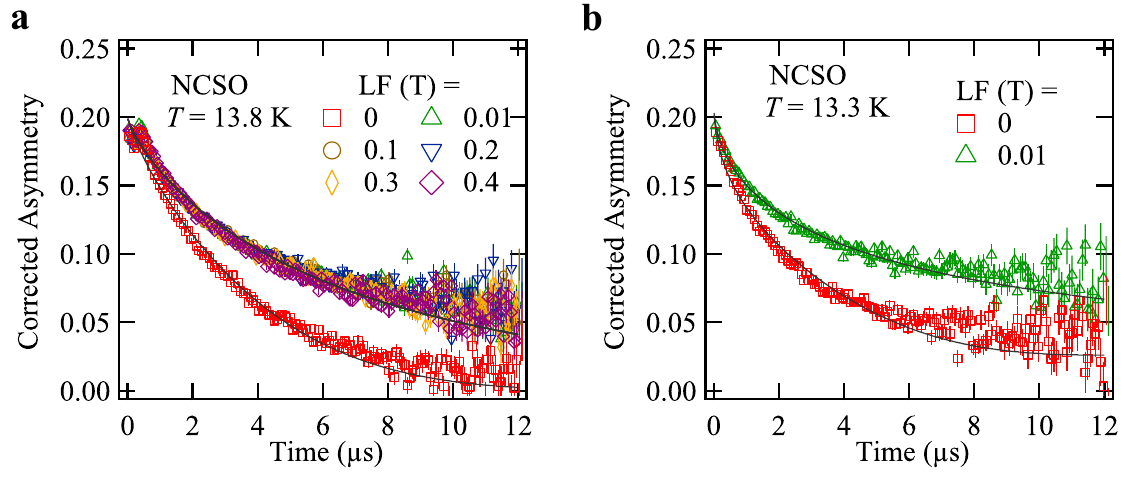}
\caption{$\mu$SR spectra of  NCSO measured using the helium-4  cryostat (a) and the helium-3  cryostat (b). The ZF spectra clearly shows that the background level for helium-4 cryostat is negligible and that for helium-3 cryostat is about 0.025. All the $\mu$SR data of NCSO discussed in the paper were measured by helium-3 cryostat except for that in (a).}
\label{figS3}
\end{figure}

\end{document}